\documentclass[fleqn,usenatbib]{mnras}

\usepackage[T1]{fontenc}
\usepackage{url}
\DeclareRobustCommand{\VAN}[3]{#2}
\let\VANthebibliography\thebibliography
\def\thebibliography{\DeclareRobustCommand{\VAN}[3]{##3}\VANthebibliography}

\usepackage{graphicx}	
\usepackage{subfig}
\usepackage{amsmath}	
\usepackage{amssymb}	
\usepackage{multirow}
\usepackage{booktabs}

\usepackage{ulem}

\title[]{Estimating Stellar Parameters from LAMOST Low-resolution Spectra}

\author[Li et al.]{
 Xiangru Li$^{1}$\thanks{E-mail: xiangru.li@gmail.com},
 and BoYu Lin$^{1}$
\\
$^{1}$School of Computer Science, South China Normal University, No. 55 West of Yat-sen Avenue, Guangzhou 510631, China\\
}

\date{Accepted XXX. Received YYY; in original form ZZZ}

\pubyear{2023}

\begin{document}
\label{firstpage}
\pagerange{\pageref{firstpage}--\pageref{lastpage}}
\maketitle

\begin{abstract}
The Large Sky Area Multi-Object Fiber Spectroscopic Telescope (LAMOST) has acquired tens of millions of low-resolution spectra of stars. This paper investigated the parameter estimation problem for these spectra. To this end, we proposed a deep learning model StarGRU network (StarGRUNet). This network was further applied to estimate the stellar atmospheric physical parameters and 13 elemental abundances from LAMOST low-resolution spectra. On the spectra with signal-to-noise ratios greater than or equal to $5$, the estimation precisions are $94$ K and $0.16$ dex on $T_\texttt{eff}$ and $\log \ g$ respectively, $0.07$ dex to $0.10$ dex on [C/H], [Mg/H], [Al/H], [Si/H], [Ca/H], [Ni/H] and [Fe/H], and  $0.10$ dex to $0.16$ dex on [O/H], [S/H], [K/H], [Ti/H] and [Mn/H],  and $0.18$ dex and $0.22$ dex on [N/H] and [Cr/H] respectively. The model shows advantages over available models and high consistency with high-resolution surveys. We released the estimated catalog computed from about 8.21 million low-resolution spectra in LAMOST DR8, code, trained model, and experimental data for astronomical science exploration and data processing algorithm research respectively.
\end{abstract}


\section{Introduction}\label{Sec:Intro}
In recent years, a series of large-scale sky survey programs have been conducted to acquire massive spectra of stars, such as the Apache Point Observatory Galactic Evolution Experiment (APOGEE) \citep{2010_APOGEE}, the Galactic Archaeology with HERMES Survey (GALAH) \citep{2015_GALAH}, the Large Sky Area Multi-Object Fibre Spectroscopic Telescope (LAMOST) Experiment for Galactic Understanding and Exploration (LEGUE) \citep{2012_LEGUE,2012_Zhao_LAMOST}, The Gaia-ESO Public Spectroscopic Survey (Gaia-ESO) \citep{2012_Gaia-ESO}, the Sloan Extension for Galactic Understanding and Exploration (SEGUE) \citep{2009_SEGUE}, the RAdial Velocity Experiment (RAVE) \citep{2008_Rave} and so on. Stellar spectra contain rich celestial information, such as the motions of the objects, atmospheric physical parameters, elemental abundances, etc. Stellar spectra information can be used in exploring stellar evolution, galaxy dynamics, etc. Therefore, the estimation of stellar atmospheric physical parameters and elemental abundances from spectra is vital in large-scale spectroscopic surveys.

Large-scale low-resolution and medium-resolution surveys are typically distinguished by its massive amount, much data with relatively low signal-to-noise ratios, and an extensive range of data quality. These difficulties challenge the computational efficiency of traditional spectral parameter estimation methods and their robustness to spectral quality. Therefore, spectral parameter estimation research based on machine learning has attracted much attention \citep{Li_2015, Xiang_2016, 2018_GPR,Zhang_2020, Xiang_2021}. The basic idea of such methods is to represent the parameter estimation problem as a mapping from spectral feature information to the parameters under being estimated. The model parameters for this mapping are determined by calculating a batch of empirical data. The parameters of each of these observed spectra are known. These parameters are usually defined based on high-quality, high-resolution spectra with the equivalent widths (EWs) method or the calculation of chemical elemental absorption lines \citep{Paula2019}.

 The traditional machine learning methods for estimating stellar spectral parameters usually consist of two key procedures: feature extraction and mapping learning. The feature extraction procedure learns an appropriate representation for stellar spectra. The spectral feature representation not only determines the interpretability and accuracy limits of the parameter estimation model, but also affects the learning difficulty of the mapping relationship \citep{Li2014, Li_2015}.
The mapping learning procedure provides the mapping relationship from the spectral information to the parameters to be estimated. Typical feature extraction methods are wavelet analysis, and wavelet packet decomposition \citep{Li_2015}, auto-encoder neural network \citep{Yang2015}, Least Absolute Shrinkage and Selection Operator (LASSO) \citep{Li2014}, principal component analysis (PCA) \citep{2018_GPR}, kernel-based principal component analysis (KPCA) \citep{Xiang_2016}, etc. The commonly used machine learning methods in stellar spectral parameter estimation are support vector machines \citep{Li2014,Zhang_2020}, linear regression \citep{Li_2015}, Gaussian process regression \citep{2018_GPR}, and neural networks \citep{Li2014}. The limitation of the traditional machine learning stellar spectral parameter estimation scheme is that the feature learning and mapping learning of the spectra are performed as two separate procedures. This characteristics results in some difficulties in designing this kind scheme and some potential improvements on parameter estimation performance.

With the advent of artificial intelligence and the big data era, deep learning methods have become the dominant methods for estimating parameters from stellar spectra, such as StarNet\citep{2018_StarNet, Zhang_2019}, AstroNN \citep{2018_AstroNN}, SPCANet\citep{ 2020_SPCANet}, and so on. These methods combine feature learning and mapping learning into a single procedure by utilizing neural networks. The procedure combination simplifies the designing of parameter estimation scheme and improves prediction performance. Therefore, neural networks promote the research and application of spectral parameter estimation of stars.

This paper investigates the problem of estimating the atmospheric physical parameters of stars and elemental abundances from low-resolution spectra of the Large Sky Area Multi-Object Fiber Spectroscopic Telescope (LAMOST). LAMOST,  also referred to as the Guo Shoujing Telescope, is at the Xinglong National Astronomical Observatory in Hebei, China. It is a special reflecting Schmidt telescope with {$4,000$} optical fibers on the focal plane. This telescope can simultaneously observe up to {$4,000$} targets in a view of a 20-square-degree field. Since 2015, LAMOST has released several versions of data from DR1 to DR8. Among them, LAMOST DR8 is the latest version of LAMOST data. The LAMOST DR8 consists of $11,214,076$ low-resolution stellar spectra covering a wavelength range of $3690$\AA-$9100$\AA, with a resolution of about 1800 at $5,500$\AA.

To estimate the parameters from LAMOST low-resolution stellar spectra, a series of studies have been carried out. These studies have also experienced the explorations from traditional machine learning schemes to deep learning solutions. Some representative studies based on traditional machine learning schemes are KPCA \citep{Xiang_2016}, The Cannon \citep{Ting_2017, Ho_2017}, SLAM \citep{Zhang_2020}, SCDD \citep{Xiang_2021} and LASSO-MLPNet \citep{ LiXiangruMNRAS2022, LiXiangruWangZhuRAA2022}. Some typical investigations of deep learning methods are GSN \citep{Rui_2019}, StarNet \citep{Zhang_2019}, DD-Payne \citep{Xiang_2019}, HotPayne \citep{Xiang_2022}, astroNN \citep{2022_Zhuohan} and Coord-DenseNet \citep{Cai_2023}). With the increasing of data volume and the development of artificial intelligence methods, deep learning methods have been more and more widely applied to estimate stellar parameters from low-resolution spectra.
However, among the methods for estimating the stellar parameters from the low-resolution spectra of LAMOST DR8, \citet{2022_Wang} focused only on the spectra with higher signal-to-noise ratio ({$S/N_{LAMOST} > 80$} and {$S/N_{APOGEE} > 70$}), \citet{LiXiangruWangZhuRAA2022,LiXiangruMNRAS2022} estimated the {$T_\texttt{eff}$}, {$\log~g$} and [Fe/H] from the spectra repectively with {$20 \le S/N_{LAMOST} \le 30$} and {$5 \le S/N_{LAMOST} \le 80$}. These constraints lead to a very limited number of samples in the reference set.
On the other hand, \citet{2022_Zhuohan} only estimated stellar parameters for giant stars, \citet{Cai_2023} only predicted the lithium abundance for some of the giant star spectra. Therefore, they only estimated the stellar parameters from a relatively small number of a small variety of parameters. Therefore, our study covers a broader range of spectral signal-to-noise ratios ({$S/N_{LAMOST}\geq5$}), estimates a wider variety of parameters ({$16$}), and includes a more significant number of stellar spectra (about {$8.21$} million).

\begin{figure*}
  \centering
  \includegraphics[width=0.8\textwidth]{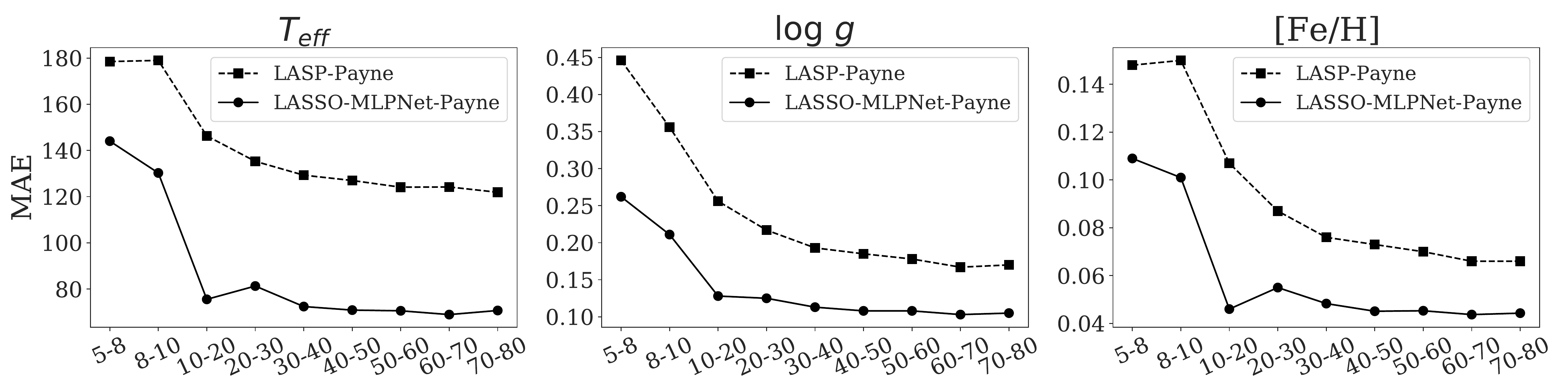}
  \includegraphics[width=0.8\textwidth]{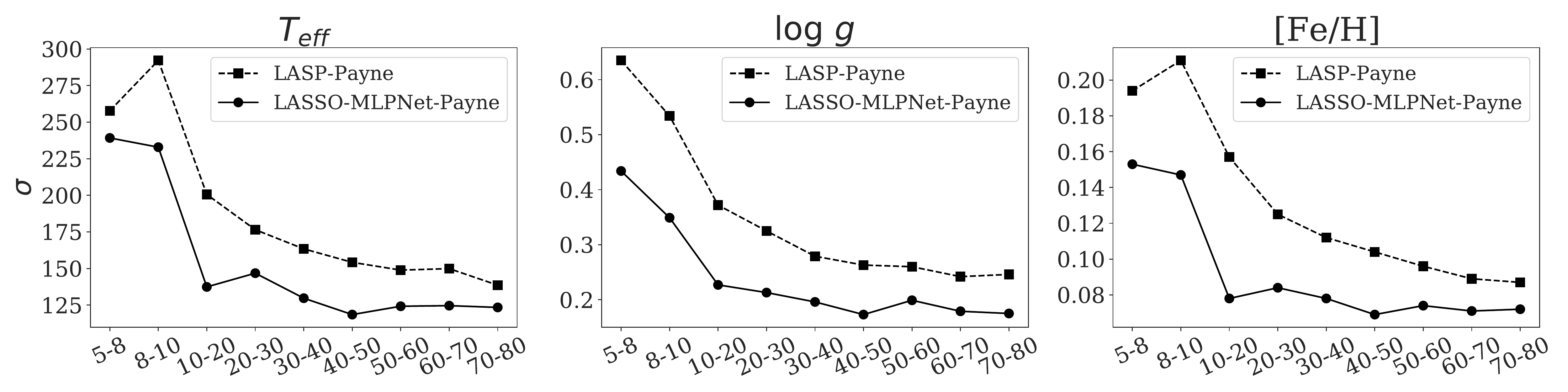}
  \caption{Parameter estimation situations of LAMOST low-resolution spectra: the dependencies of the MAE error of the estimations from LASP \citep{2015_LASP} and LASSO MLPNet \citep{LiXiangruMNRAS2022} on the SNR. MAE: mean absolute of error. $\sigma$: standard deviation of error. LASP: LAMOST Stellar Parameter Pipeline.}
  \label{fig_lasso:widgets}
\end{figure*}

To determine the stellar parameters (effective temperature, surface gravity, and metal abundance) for the vast amount of LAMOST spectral data,  researchers developed the LAMOST Stellar Parameter Pipelines (LASP) \citep{2015_LASP}.
LASP provides parameter estimation results for LAMOST DR8 low-resolution spectra using the ELODIE spectral library as templates and a $\chi^2$ minimization method based on the ULySS procedure \citep{2011_Wu}. However, our preliminary study shows that the precision of the LASP estimation results decreases rapidly with the decline of the signal-to-noise ratio (SNR) of spectra. In the case of $5\leq S/N_g < 8$, $8\leq S/N_g < 10$, $10\leq S/N_g <  20$, and $20\leq S/N_g < 30$, the mean absolute of error (MAE) of LASP are $178.5$ K, $179.0$ K, $146.3$ K, and $135.3$ K on $T_\texttt{eff}$, $0.446$ dex, $0.356$ dex, $0.256$ dex, and $0.217$ dex on $\log\ g$, $0.148$ dex, $0.150$ dex, $0.107$ dex, and $0.087$ dex on [Fe/H], while the standard deviation of error ($\sigma$) are $257.8$ K, $292.2$ K, $200.6$ K, and $176.5$ K on $T_\texttt{ eff}$, $0.635$ dex, $0.534$ dex, $0.372$ dex, and $0.325$ dex on $\log~g$, $0.194$ dex, $0.211$ dex, $0.157$ dex, and $0.125$ dex on [Fe/H]. Therefore, \citet{LiXiangruWangZhuRAA2022,LiXiangruMNRAS2022} correspondingly conducted some investigations and improved the precision of the parameter estimates compared to LASP (Figure \ref{fig_lasso:widgets}). However, \citet{LiXiangruWangZhuRAA2022,LiXiangruMNRAS2022} and LASP are limited to estimating three stellar atmospheric physical parameters, $T_\texttt{eff}$, $\log~g$ and [Fe/H]. Therefore, this paper focused on further improving the precision of stellar atmospheric physical parameter estimation while also investigating the measurement of 13 more elemental abundances ([C/H], [Mg/H], [Al/H], [Si/H], [Ca/H], [N/H], [O/H], [S/H], [Ti/H], [Cr/H], [Mn/H], [Ni/H], and [K/H]).

The implementation code of the proposed neural networks in this paper are done in Tensorflow. The full project and its documentation are available at \url{http://doi.org/10.12149/101216} after acceptance for publication. This project includes the estimated catalog computed from about $8.21$ million low-resolution spectra in LAMOST DR8, code, trained models, and experimental data for astronomical science exploration and data processing algorithm research, respectively. The project documentation will provide a detailed description of the overall project architecture.

The remainder of this paper is organized as following. Section \ref{Sec:Data_Preprocessing} presents the data used in this paper; Section \ref{Sec:Methods} describes the proposed methodology, its evaluation, and model uncertainty analysis; Section \ref{Sec:Applications} shows our application results on approximately $8.21$ million low-resolution spectra from LAMOST; Section \ref{Sec:Conclusion} gives some conclusions.

\section{Reference Datasets and Their Preprocessing}\label{Sec:Data_Preprocessing}

The proposed scheme in this paper is one of machine learning methods. This kind method need a reference data set (referred to as a reference set). The reference set is used for learning the model parameters in the mapping from the spectral information to the stellar parameters to be estimated. Therefore, the reference set is a knowledge carrier for the stellar parameter estimation problem, consists of the observed spectra and their stellar atmospheric physical parameters and elemental abundances. The observed spectra in the reference set are obtained from the LAMOST DR8 low-resolution spectral library. The stellar atmospheric parameters and elemental abundances of the observed spectra are obtained from the APOGEE DR17 catalog. The spectral parameters estimated in this work include the effective temperature $T_\texttt{eff}$, surface gravity $\log\ g$, and 14 elemental abundances [X/H] (X refers to C, N, O, Mg, Al, Si, S, K, Ca, Ti, Cr, Mn, Fe, Ni).

\subsection{APOGEE and APOGEE DR17 catalog}\label{Sec:Data_Preprocessing:APOGEE}

LAMOST spectra have a low resolution, and the SNR of a large fraction of them is below $30$. These characteristics result in a considerable improvement space for the estimation precision of the LASP estimation from LAMOST spectra. Moreover, LASP does not give abundance estimates for elements other than [Fe/H]. One possible solution is to transfer parameter information from other high-resolution and high-quality survey spectral libraries to the LAMOST spectral library based on the spectra from common sources.

The Apache Point Observatory Galactic Evolution Experiment (APOGEE) \citep{2010_APOGEE} is a high-resolution infrared sky survey based on the Sloan telescope, with a band coverage from $1.51\mu m$ to $1.70\mu m$. ASPCAP (The APOGEE Stellar Parameter and Chemical Abundances Pipeline) gives estimates of ${T}_\texttt{eff}$, $\log \ g$, and chemical elemental abundances for APOGEE spectra. APOGEE DR17 catalog published the atmospheric parameters (${T}_\texttt{eff}$, $\log \ g $, [Fe/H]) and elemental abundances for 475,144 stars.  The ranges of the stellar atmospheric parameters in the APOGEE DR17 catalog are $[3500,7000]$ K for ${T}_\texttt{eff}$, $[-0.5,5]$ dex for $\log \ g $, and $[-2.0,0.5]$ dex for [Fe/H].

Therefore, this paper builds a reference dataset by cross-matching the APOGEE DR17 catalog and the LAMOST DR8 low-resolution spectral library. Each sample in this dataset consists of one LAMOST low-resolution spectrum and the estimations from the common source observation in APOGEE DR17. The final reference set consists of $240,448$ observed spectra and their corresponding stellar parameters. The spectral parameters explored in this paper include the stellar atmospheric physical parameters $T_\texttt{eff}$, $log~g$, [Fe/H], and $13$ elemental abundances [X/H], where X refers to C, N, O, Mg, Al, Si, S, K, Ca, Ti, Cr, Mn, and Ni.

It is shown that there is a difference in the effective features of parameter estimation between low S/N spectra and high S/N spectra.  To increase the parameter estimation performance by detecting the spectral features adaptive to the spectral quality, therefore, the reference set are further divided into two subsets $S^{lS/N}$ and $S^{hS/N}$ based on the signal-to-noise ratio criterion $5 \le S/N_g \le 50$ and $S/N_g > 50$. The sample sizes of these two reference subsets are $96,200$ and $144,248$, respectively. For the reference set $S^{lS/N}$, we randomly divide it into a training set $S^{lS/N}_{tr}$, a validation set $S^{lS/N}_{val}$ and a test set $S^{lS/N}_{te}$ in the ratio of 7:1:2. The sample sizes of $S^{lS/N}_{tr}$, $S^{lS/N}_{val}$ and $S^{lS/N}_{te}$ are $67,340$, $9,620$, and $19,240$, respectively.
These three reference sets were used respectively for training, hyperparameter selection, and performance evaluation for the parameter estimation model used on the spectra with low $S/N_g$.
Similarly, we randomly divide the reference set $S^{hS/N}$ into three subsets $S^{hS/N}_{tr}$, $S^{hS/N}_{val}$ and $S^{hS/N}_{te}$, which are used for training, hyperparameter selection and performance evaluation for the parameter estimation model used on the spectra with high $S/N_g$. The sample numbers of $S^{hS/N}_{tr}$, $S^{hS/N}_{val}$ and $S^{hS/N}_{te}$ are $100,973$, $14,425$ and $28,850$, respectively.

\subsection{Data preprocessing}\label{Sec:Data_Preprocessing:Processing}

The observed spectra are negatively affected by many factors, such as redshift, noise, and skylight. These factors can decrease the precision and stability of parameter estimation, and the demand for more reference data \citep{2022_RRNet}. Therefore, the stellar spectral data must be preprocessed before input into the parameter estimation model. The specific preprocessing steps are as follows.

\textbf{Wavelength correction.} We used radial velocity (RV) for wavelength correction to move each spectrum to its rest frame:
\begin{equation}
    \lambda'=\frac{\lambda}{1+\text{RV}/c},
\end{equation}
where $\lambda'$,  $\lambda$ , $c$ and RV respectively denote the corrected wavelength, the original wavelength, the speed of light, and the radial velocity. In this paper, the wavelength correction is performed using the radial velocity estimates given by the official LAMOST stellar parameter estimation pipeline (LASP).

\textbf{Linear interpolation resampling.} We utilized the maximum common wavelength range $[3841 \text{\AA},5699 \text{\AA}]$ and $[5901 \text{\AA},8798 \text{\AA}]$ respectively for the blue end and red end of all spectra. Based on the common wavelength range, we resampled each spectrum using a linear interpolation method with a resampling step size $0.0001$dex in logarithmic space.

\textbf{Denoising.} The observed spectra are usually contaminated with bad pixels and impulse noise, which can negatively affect the mapping learning of the model. Therefore, the observed spectra need to be denoised. To this end, we used the median filtering method to reduce the spectral noise. The size of the filtering window is 3 pixels.

\begin{figure*}
\centering
\subfloat[A LAMOST spectrum.]{
	\includegraphics[width=0.8\textwidth]{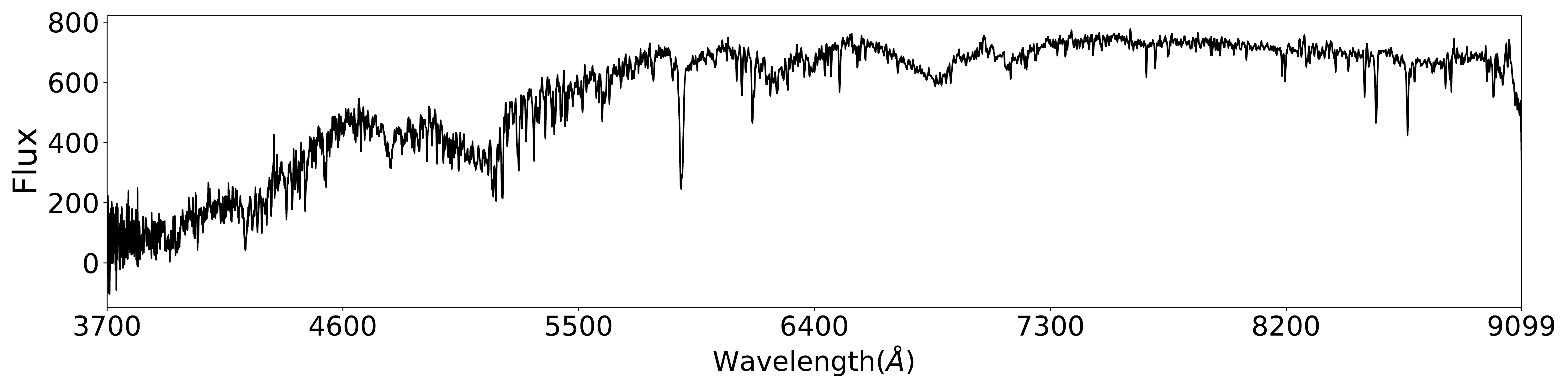}
	\label{sf1}
	}\\
\subfloat[The blue-end spectrum and the red-end spectrum after wavelength correction, linear interpolation resampling, and denoising. The dashed line indicates the estimated continuum.]{
	\includegraphics[width=0.8\textwidth]{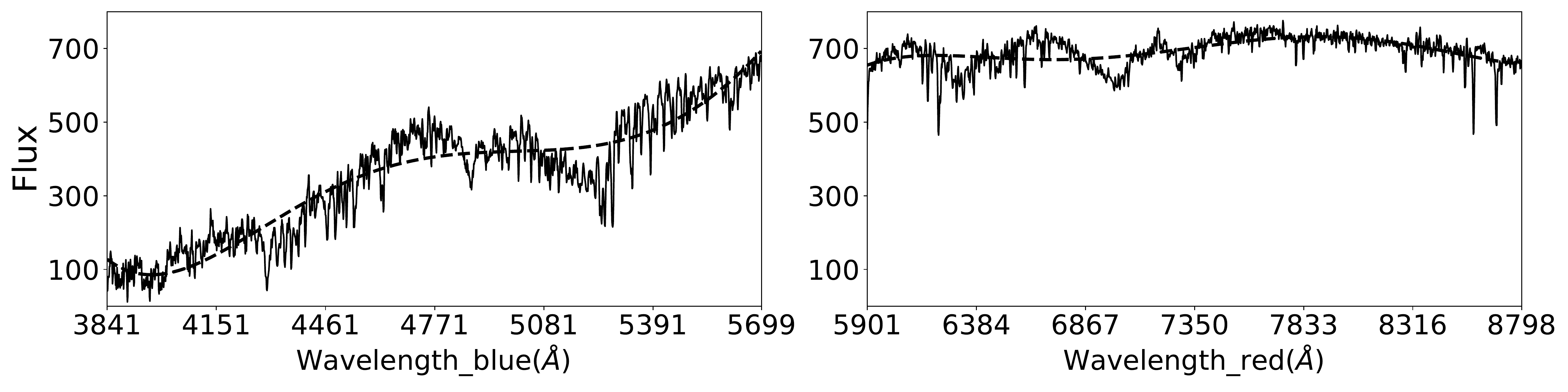}
	\label{sf2}
	}\\
	\subfloat[The blue-end spectrum and the red-end spectrum after continuum normalization.]{
	\includegraphics[width=0.8\textwidth]{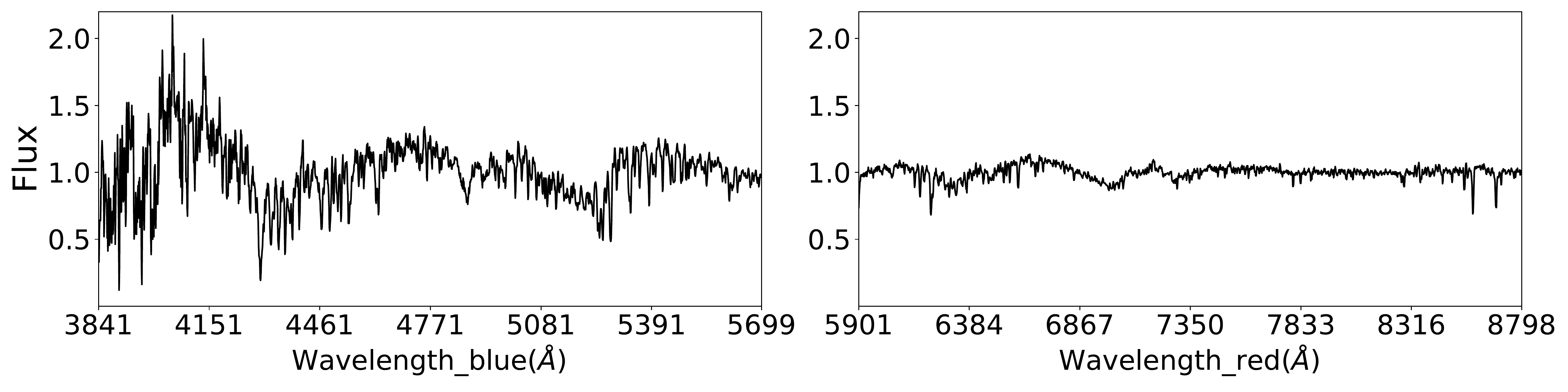}
	\label{sf3}
	}\\
	\subfloat[The blue-end spectrum and the red-end spectrum after secondary denoising and spectrum-wise normalization.]{
	\includegraphics[width=0.8\textwidth]{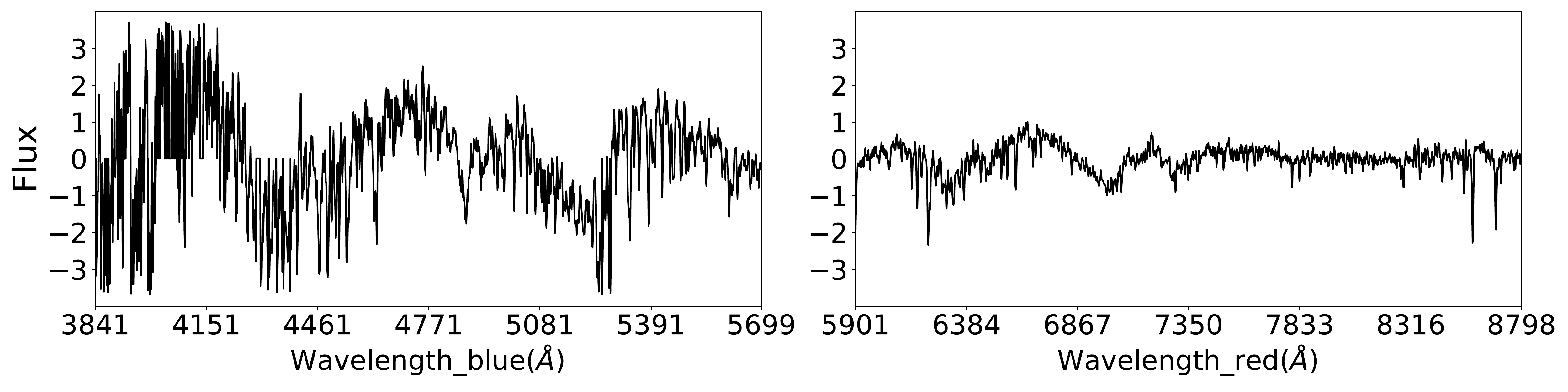}
	\label{sf4}
	}
\caption{A LAMOST DR8 low-resolution spectrum (spec-55863-M31\_011N40\_B1\_sp08-198) and its pre-processing results. The horizontal and vertical coordinates characterize the wavelength and flux, respectively.}
\label{fig2:widgets}
\end{figure*}

\textbf{Continuum normalization.} Since the spectrophotometric correction applied to these spectra is only an approximation, the observed fluxes at different wavelengths are not accurate in an absolute sense. Therefore, continuum normalization \citep{Fiorentin_2007,2020_SPCANet,LiXiangruMNRAS2022} is required prior to parameter estimation. The basic step of continuum normalization is to estimate the continuum of every spectrum by curve fitting first. This estimated continuum is referred to as a pseudo-continuum. Then, each pixel of a spectrum is divided by the flux of the corresponding pseudo-continuum. The pseudo-continuum is an estimation of the trend in the dependencies of the spectral fluxes on wavelength (Figure \ref{sf2}). The continuum is generally estimated by a polynomial fitting method \citep{Fiorentin_2007,2020_SPCANet, LiXiangruMNRAS2022}. In this paper, the pseudo-continuums are estimated separately for the blue-end and red-end spectra using a 5th-order polynomial fitting.

\textbf{Secondary denoising and spectrum-wise-normalization.} After continuum normalization, there exist negative effects from some aberrant variation range on fluxes between various spectra, and some interferences from non-impulse noises. The presence of non-impulse noise reduces the sensitivity of the algorithm to weak spectral features. Therefore, each continuum-normalized spectrum $\mathbf{x}=(x_1, \cdots, x_D)^T$ is further processed as follows: in case of a flux is smaller than $\mu-3\sigma$ or larger than $\mu+3\sigma$, this flux is replaced by $\mu$; and each spectral flux $x_i$ is transformed as follows:
\begin{equation}
    z_i = \frac{x_i-\mu}{\sigma}, i = 1, \cdots, D,
\end{equation}
where $\mu = \sum\limits_{i=1}^D{x_i}/D$ and
  $$\sigma = \sqrt{\sum\limits_{i=1}^D{(x_i - \mu)^2}/D}.$$
Figure \ref{fig2:widgets} shows a spectrum and its preprocessing results.  It is shown that the spectral features are significantly enhanced after pre-processing.

\section{Stellar Spectral Parameter Estimation Method StarGRUNet and its evaluations}\label{Sec:Methods}

\subsection{StarGRUNet}

The proposed stellar spectral parameter estimation scheme is an artificial neural network (NN). The NN is a hierarchically organized computational model. More about NN can be found in \citep{Goodfellow2016,LiXiangruMNRAS2022}. The proposed NN is presented in Table \ref{table:BGANet}. Compared with the previous work \citep{LiXiangruMNRAS2022}, the BGANet model is further equipped with some Bi-GRU learning layers and a self-attention learning layer. The Bi-GRU learning is to exploit the correlation information between various wavelength subbands, and the Self-Attention learning module is to discover parameter-sensitive features of different types of spectra automatically. For more information about Bi-GRU and Self-Attention learning, please refer to \citet{NiuZhaoYang2021SelAttention}.

\begin{table}
\centering
\caption{ The BGANet network. In step 1), there is a model parameter $t$, which indicates the numbers of wavelength subbands; in step 2), there are parameters $n$ and $l_1, \cdots, l_n$, which indicate the numbers of Bi-GRU layers and the feature dimension of each Bi-GRU learning layer, respectively.}
\begin{tabular}{|c|c|}
\hline
\textbf{Steps} & Calculations                   \\ \hline
Input                & Pre-processed spectra               \\ \hline
1) & \begin{tabular}[c]{@{}c@{}}Dividing each spectrum into  $t$ subbands\\ with equal wavelength width.\end{tabular} \\ \hline
2)                   & A series of Bi-GRU learning layers  \\ \hline
3)                   & A Self-Attention learning layer     \\ \hline
4)                   & A fully connected learning layer    \\ \hline
Output               & An estimated spectral parameter \\ \hline
\end{tabular}\label{table:BGANet}
\end{table}

Due to the influence of random factors in model initialization and the learning process, the generalization ability of individual BGANet generally can be improved further. One solution is to employ an ensemble learning strategy to combine the learning results of several BGANet models. The fundamental idea of ensemble learning is to improve prediction performance by training multiple BGANet learners and exploiting their complementary capabilities. The typical methods for combining the regression prediction results of several learners are simple average, weighted average, and learning techniques.

\begin{figure}
  \centering
  \includegraphics[width=0.45\textwidth]{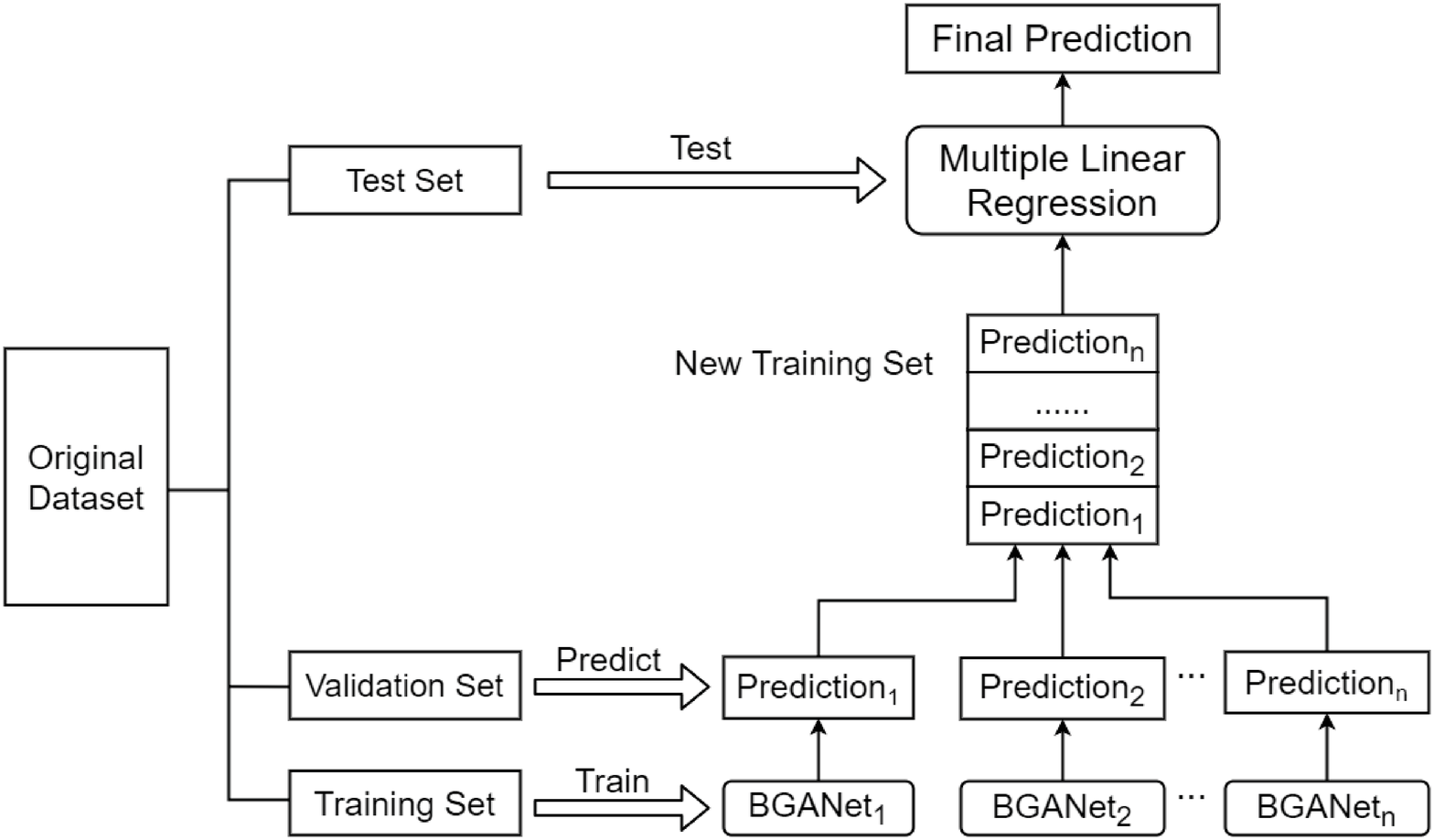}
  \caption{ The principles of the proposed StarGRUNet.}
  \label{fig52:widgets}
\end{figure}

The distinctive characteristics of the stellar spectral parameter estimation problem studied in this paper are the large size of the reference dataset and the large amount of model parameters. These characteristics require that the ensemble learning strategy should be easy to be implemented, efficient, and stable. Therefore, we adopted the Blending learning strategy--a simplified version of the Stacking learning method \citep{1992_Stacking} and formed the StarGRUNet method, whose principle can be found in Figure \ref{fig52:widgets}.

Taking the estimation of parameter $T_\texttt{eff}$ as an example, the training steps of StarGRUNet are as follows. Suppose $S_{val} = \{(\textbf{x}_i, y_i), i=1,\cdots,s\}$ is a validation set. First, for each spectrum $x_i\in S_{val}$, we estimated its $T_\texttt{eff}$ using $n$ trained BGANet models and computed a vector $\textbf{z}_i=(z_i^1, \cdots, z_i^n)^T$. Second, treat $S‘_{val} = \{(\textbf{z}_i, y_i), i=1,\cdots,s\}$ as a training set to learn the secondary learner using a multiple linear regressor. The secondary learner is to fuse the estimations from $n$ BGANet models. The models for estimating other stellar parameters can be trained similarly.

\subsection{Model Selection and Model Training}\label{Sec:Methods:ModelSelection}

Model hyperparameters can significantly affect predictive performance. There are two sets of hyperparameters in the BGANet model. The first set of hyperparameters consists of the number of wavelength subbands $t$ and the number of Bi-GRU layers $n$. In the Bi-GRU module, we index the subbands with $i=1,2,3,... ,t$ from left to right. In case of a small $t$, there is less communication between different wavelength subbands, and it is necessary to take a smaller value for $n$ to reduce the risk of overfitting. In case of a big $t$, more communication and more complex interdependencies are investigated between various wavelength subbands. Therefore, larger values of $n$ are needed to enhance the model complexity by exploiting more complex cross-band correlations and complementarities. Besides, the choice of $t$ and $n$ is theoretically related to the size of the training set. The larger the parameters $t$ and $n$, the higher the model complexity and the more training data are needed for model learning. This work suggests several configurations of $2$ or $3$ for $n$ and $5$, $10$, or $15$ for $t$ based on experimental experiences (Table \ref{tab1:widgets}).

The second set of hyperparameters is the dimensions $\{l_j, j=1, \cdots, n \}$ of the features extracted from various Bi-GRU layers, where $j$ is the index of a Bi-GRU layer. A negative correlation should be maintained between $l_j$ and $j$. A small $j$ indicates that the corresponding Bi-GRU layer is close to the input end of the BGANet, and a large $l_j$ should be set in this case to effectively extract the spectral features as possible. Similarly, a large $j$ indexes indicates that the corresponding Bi-GRU layer is close to the output end of the BGANet, and a small $l_j$ should be set to reduce noises, redundancies, and the risk of overfitting in mapping learning. In addition, the parameters $\{l_j, j=1, \cdots, n \}$ also determine the complexity of the BGANet model. A BGANet with a small $l_j$ has relatively few model parameters and a low model complexity; on the contrary, the BGANet with more model parameters is more complex.

Based on the above-mentioned principles and some experimental experiences, we selected three BGANet models with excellent prediction results on the validation set (Table \ref{tab1:widgets}). We took these models as primary learners for StarGRUNet. To estimate each spectral parameter, we built a StarGRUNet model respectively for the spectra with low SNR and high SNR.

\begin{table}
\centering
\caption{ The proposed configuration for the hyperparameters of three BGANets in the proposed StarGRUNet.}
\vspace{0.2cm}
\begin{tabular}{|c|c|c|c|c|c|}
\hline
\textbf{Model} & {$\boldsymbol n$} & {$\boldsymbol t$} & {$l_1$} & {$\boldsymbol l_2$} & {$\boldsymbol l_3$} \\ \hline
BGANet1 & 2 & 5  & 64  & 32 & ... \\ \hline
BGANet2 & 3 & 10 & 128 & 64 & 32  \\ \hline
BGANet3 & 3 & 15 & 128 & 64 & 32  \\ \hline
\end{tabular}\label{tab1:widgets}
\end{table}

\subsection{Model Evaluation}\label{Sec:Methods:Evaluation}

In this subsection, we evaluated the performance of StarGRUNet on the test set. The evaluations are conducted based on the following metrics: $\mu$ - the mean of the difference between StarGRUNet predictions and the APOGEE DR17 catalog, $\sigma$ - the standard deviation of the difference, and MAE - the mean of the absolute of difference. Among them, $\mu$ indicates the deviation or inconsistency between the prediction result and the reference. $\sigma$ measures the degree of dispersion or instability of the consistency between the prediction results and the reference. And MAE is a cumulative measure of the difference on all test samples and describes the overall inconsistency.

\begin{figure}
  \centering
  \includegraphics[width=0.49\textwidth]{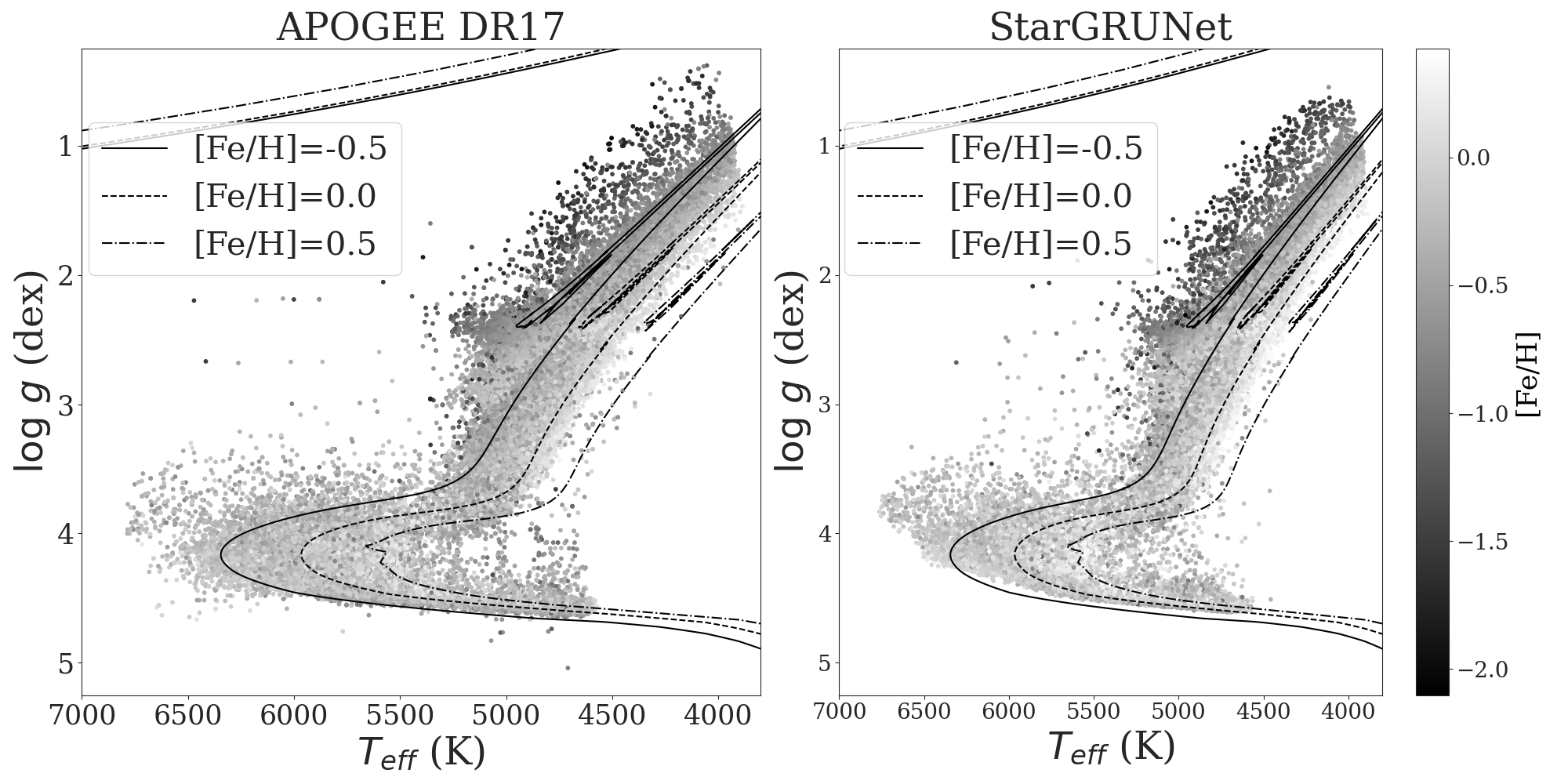}
  \caption{Comparison between the StarBRUNet predictions and APOGEE DR17 catalog on the test set. The left subplot shows the results from the APOGEE DR17 catalog, and the right subplot shows the estimated results from StarGRUNet. The colors indicate the [Fe/H] abundances. The solid line, the dashed line, and the dotted line respectively indicate three MIST stellar isochrones with stellar ages of 7 Gyr.}\label{fig8:widgets}
\end{figure}

To evaluate the performance of StarGRUNet, we compared its estimation results with the APOGEE DR17 catalog in $T_\texttt{eff}-\log\ g$ space (Figure \ref{fig8:widgets}). For easy comparison, three MIST stellar isochrones with stellar ages of $7$ Gyr were presented in this figure. It is shown that the StarGRUNet predictions not only reconstruct APOGEE DR17's $T_\texttt{eff}$ and $\log\ g$ nicely but also match the MIST stellar isochrones well. These phenomena indicate a strong consistency between the predictions of stellar atmospheric parameters from StarGRUNet and the APOGEE DR17 catalog.

\begin{figure*}
   \centering
  \includegraphics[width=0.23\textwidth]{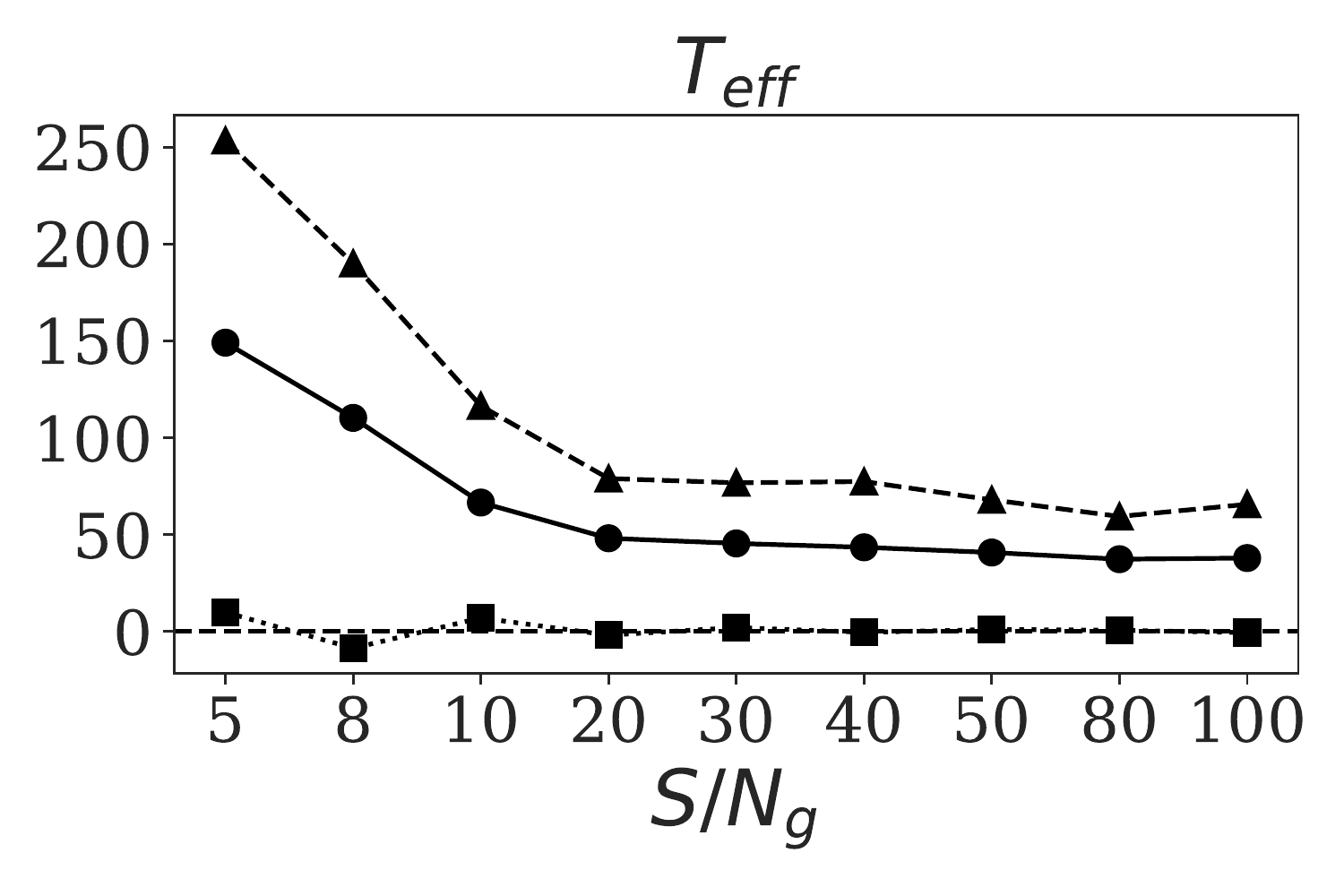}
  \includegraphics[width=0.23\textwidth]{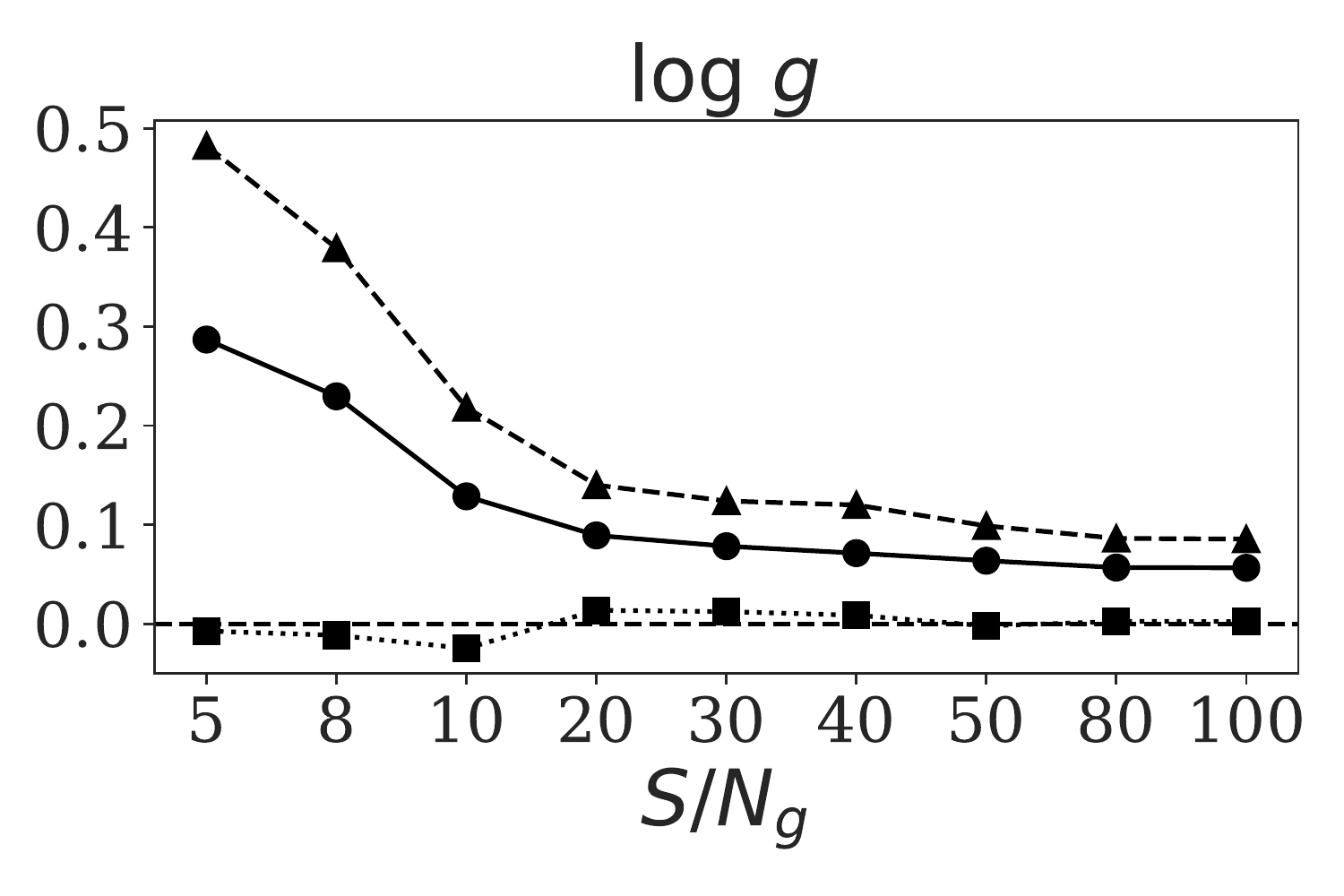}
  \includegraphics[width=0.23\textwidth]{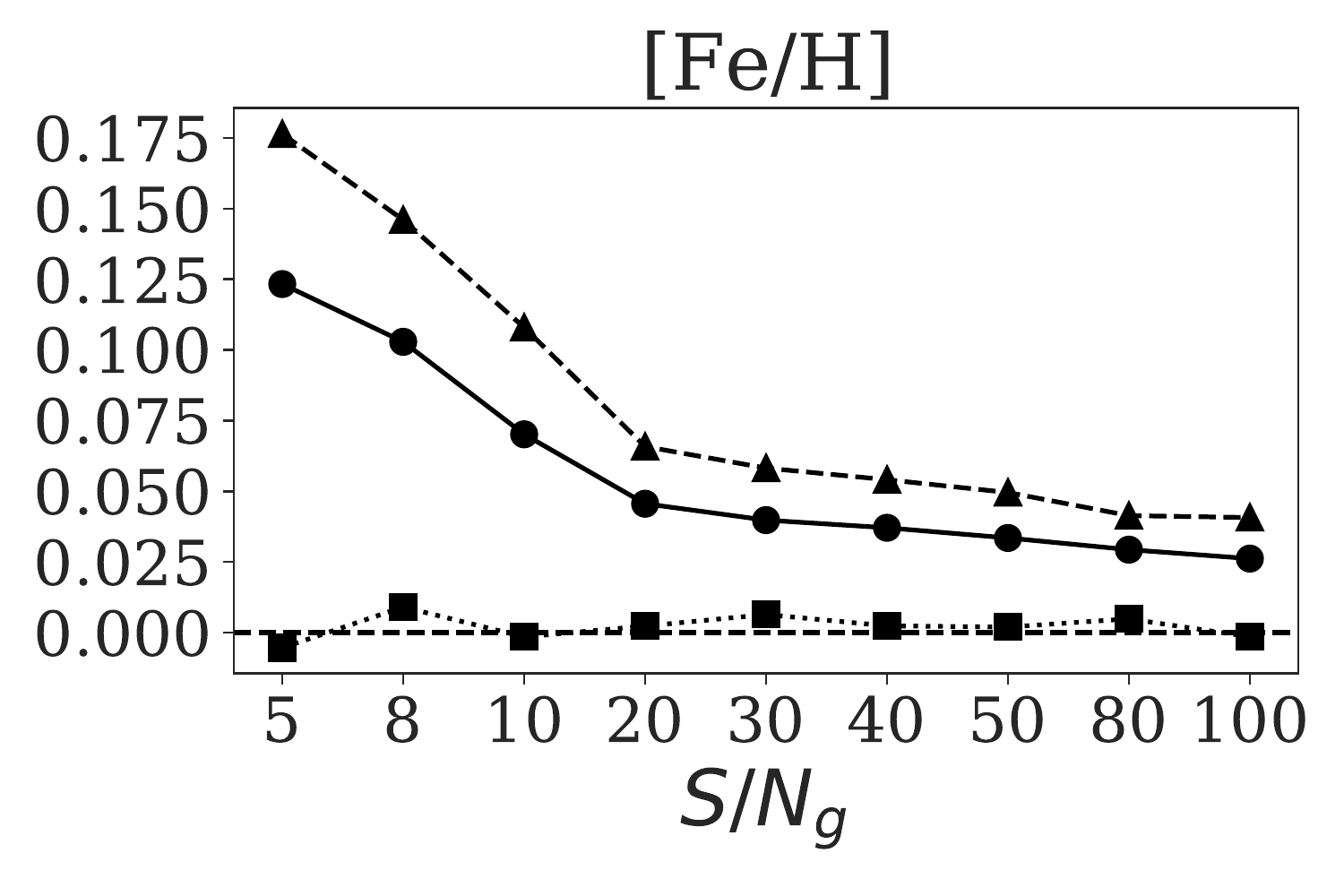}
  \includegraphics[width=0.23\textwidth]{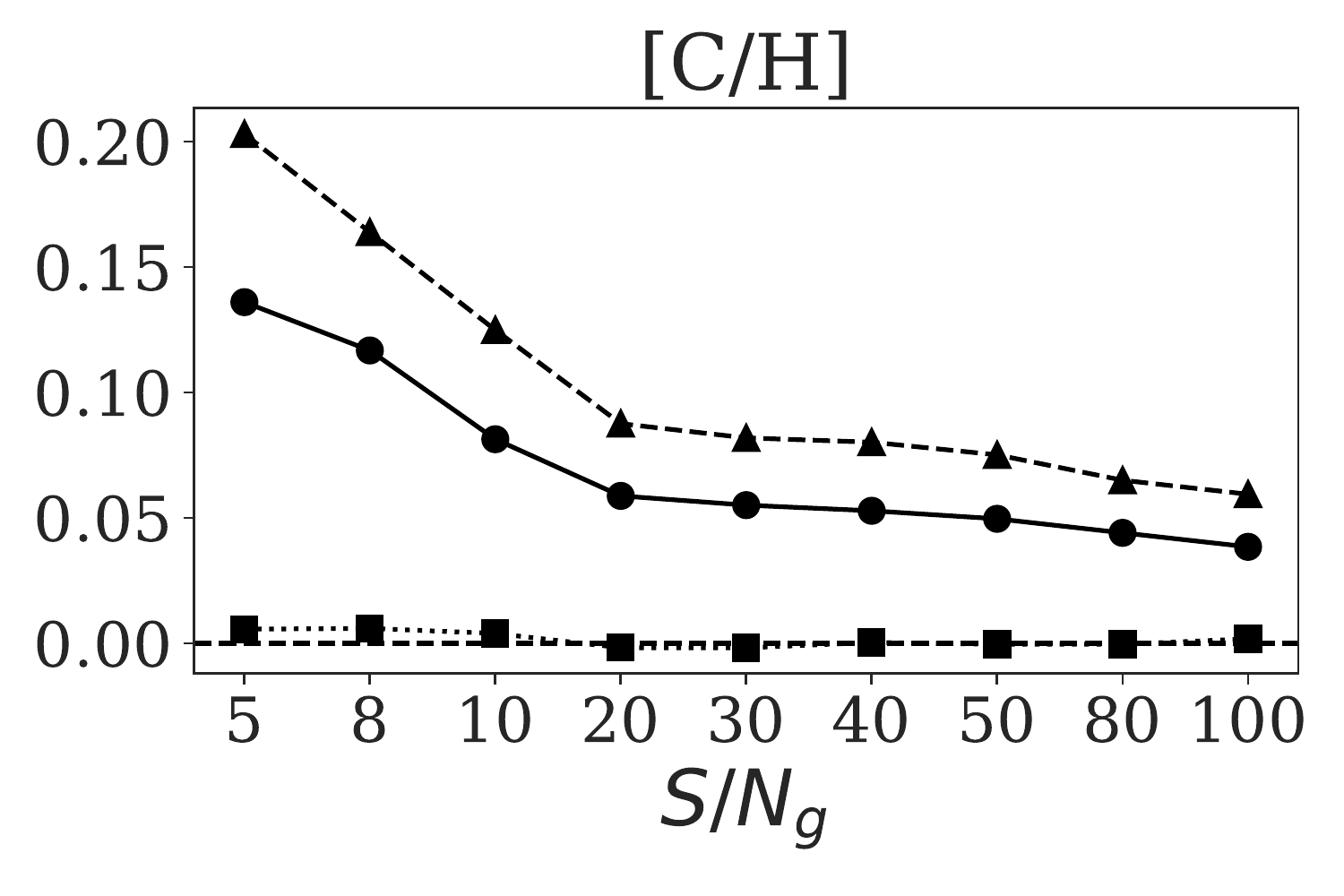}
  \includegraphics[width=0.23\textwidth]{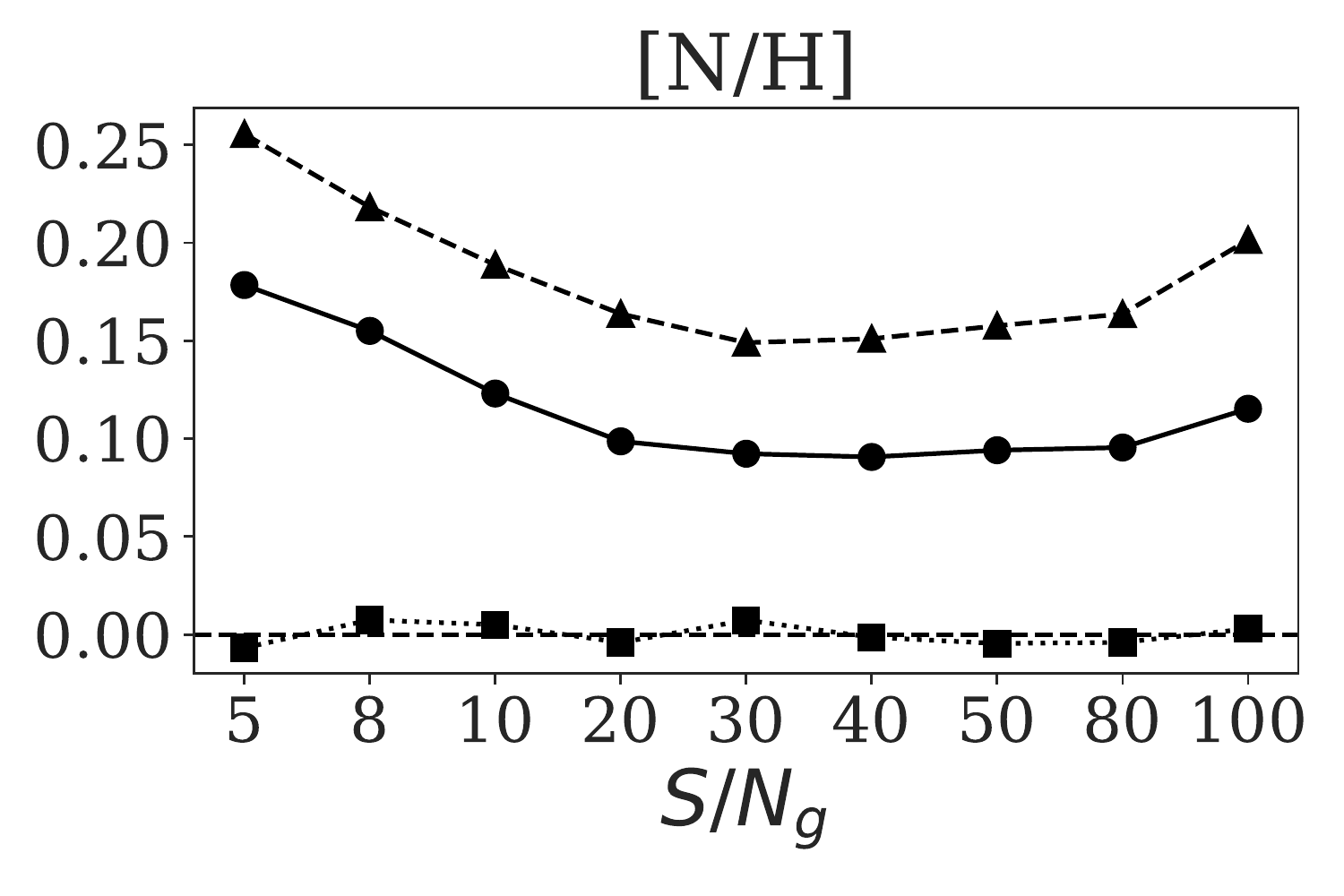}
  \includegraphics[width=0.23\textwidth]{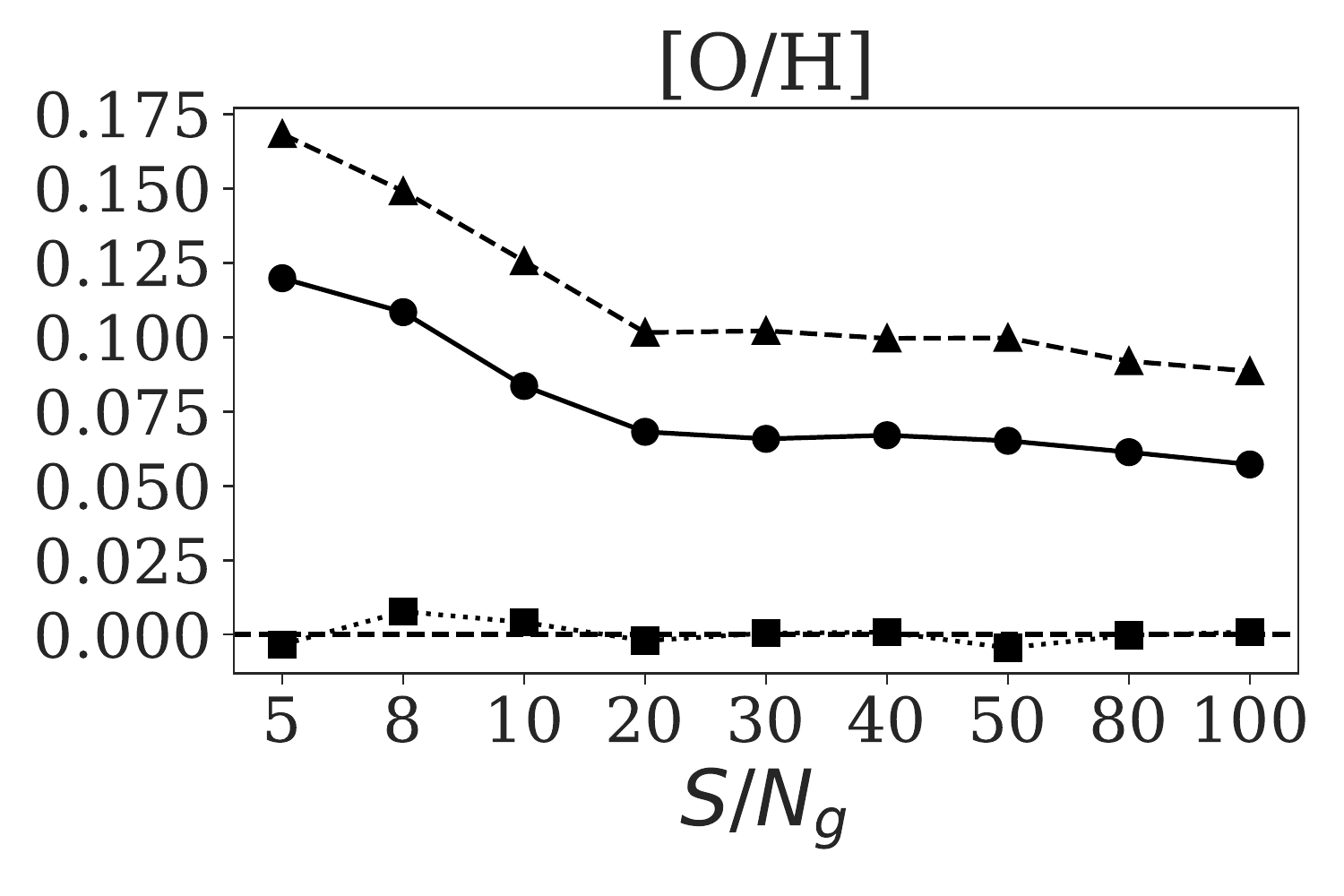}
  \includegraphics[width=0.23\textwidth]{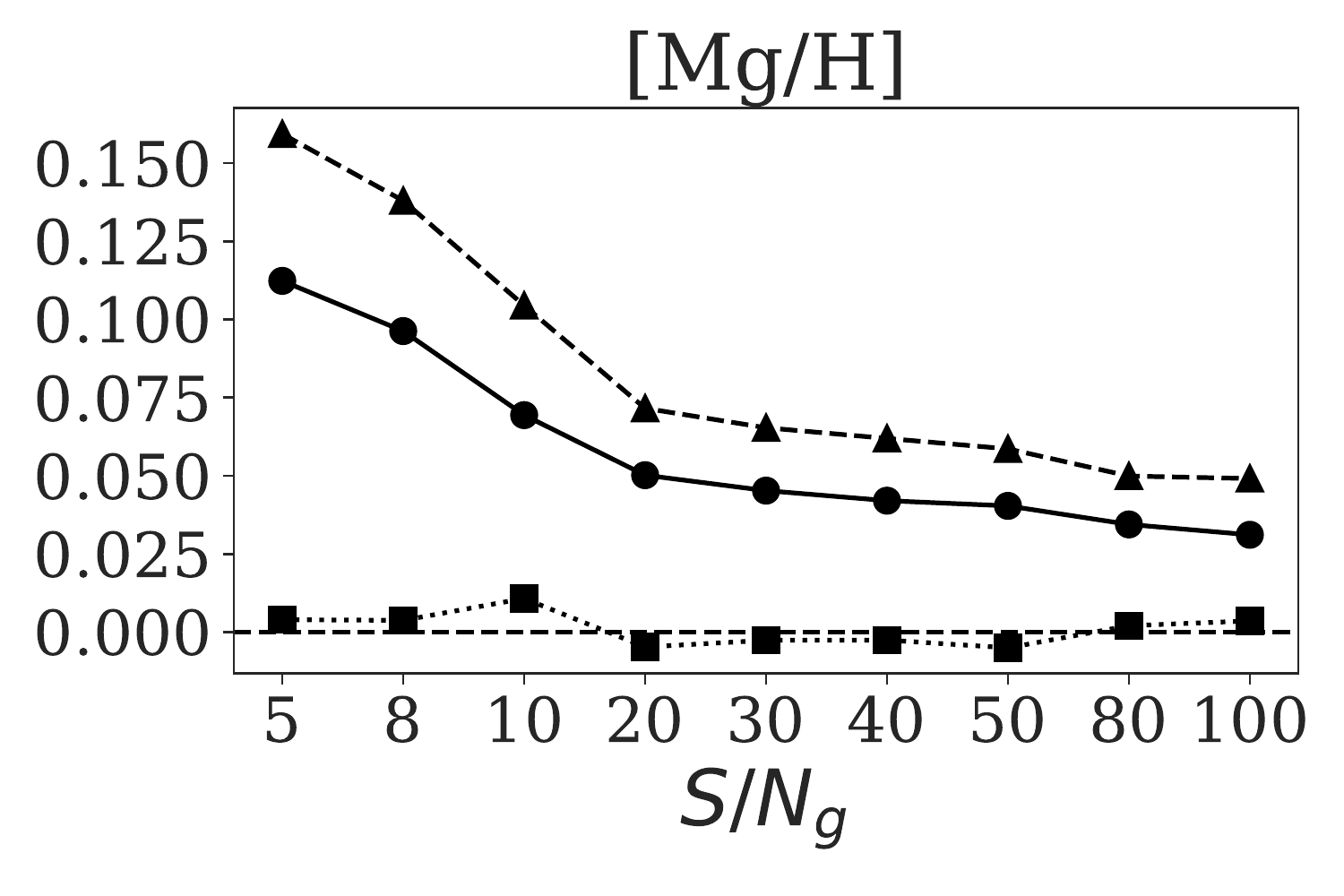}
  \includegraphics[width=0.23\textwidth]{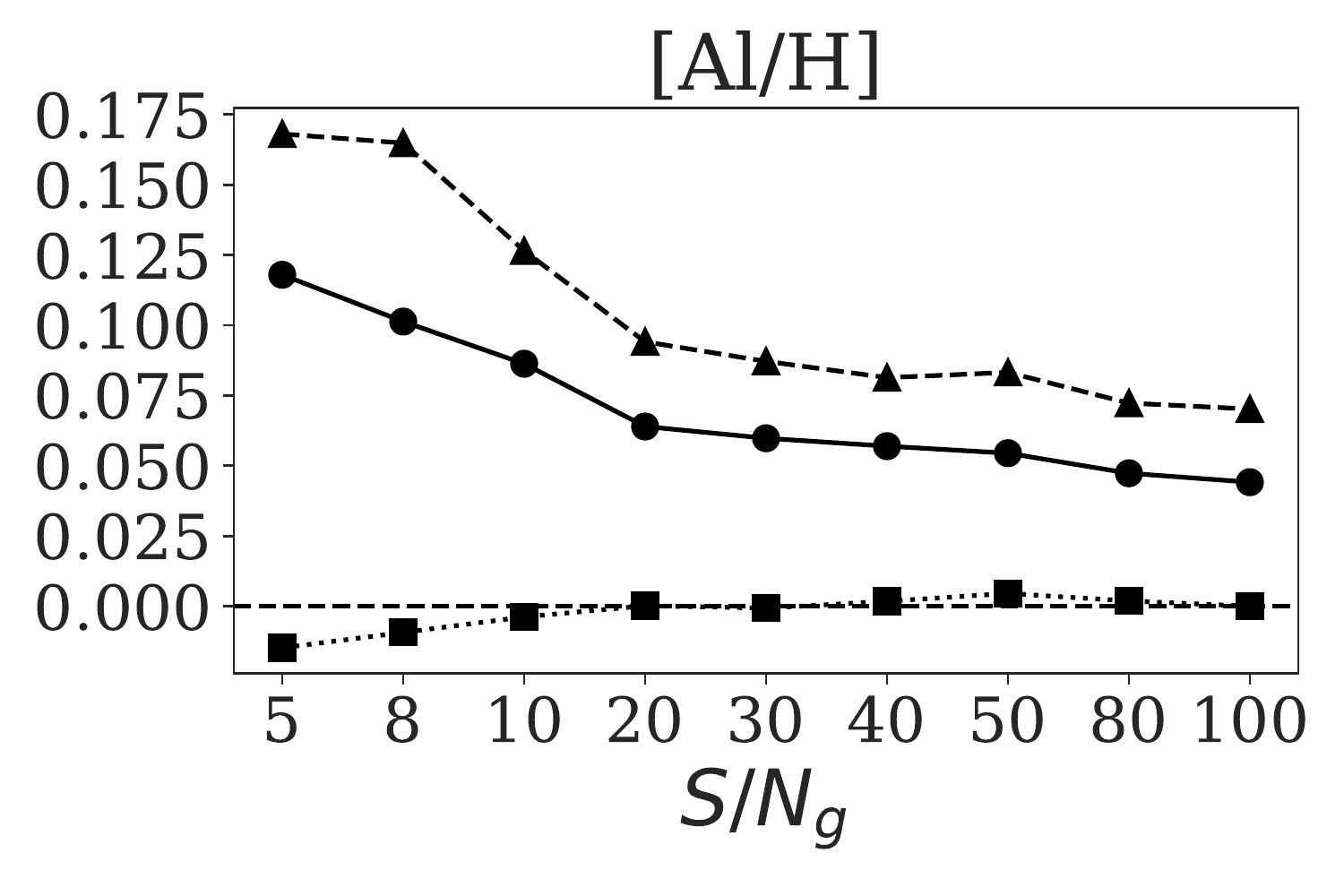}
  \includegraphics[width=0.23\textwidth]{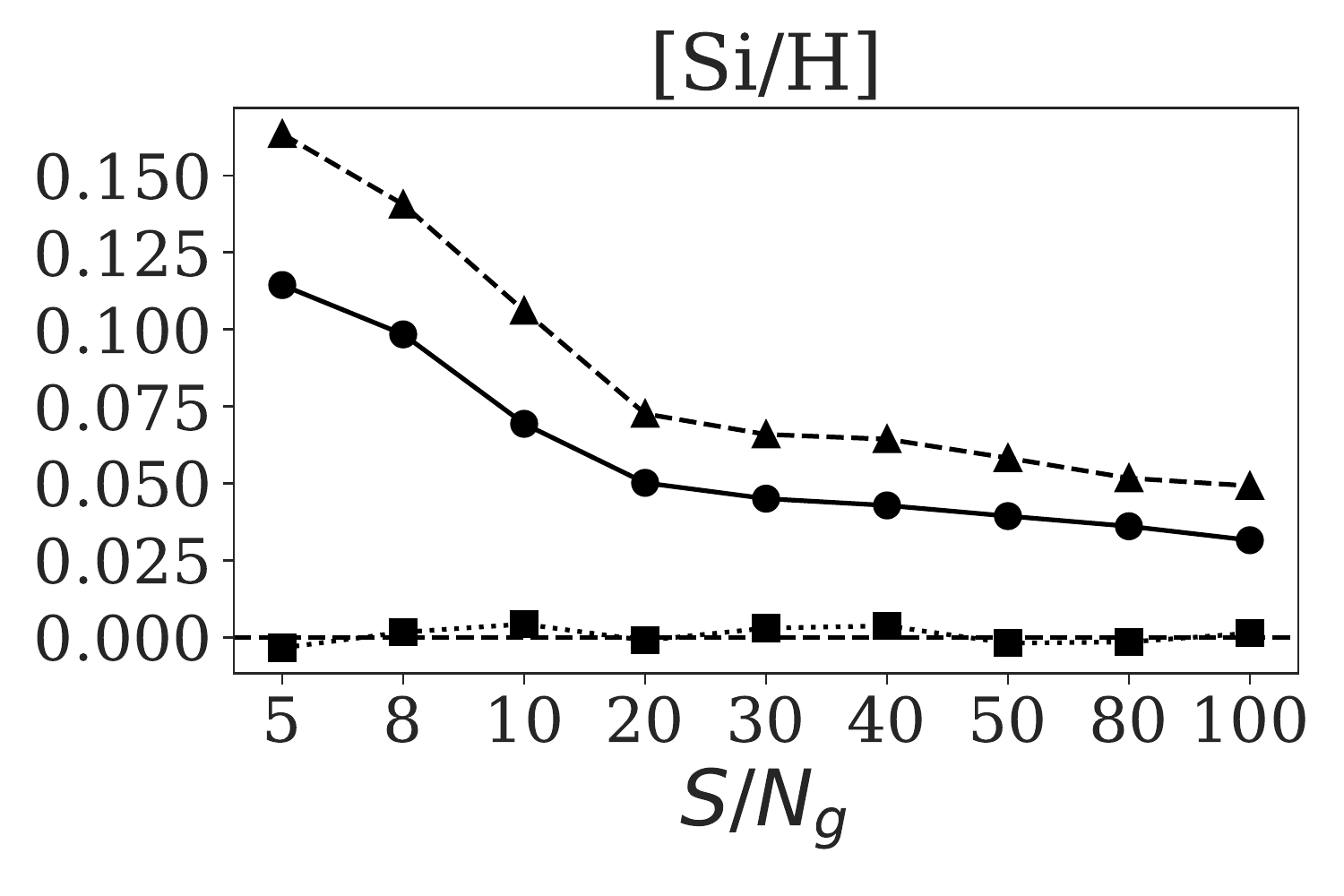}
  \includegraphics[width=0.23\textwidth]{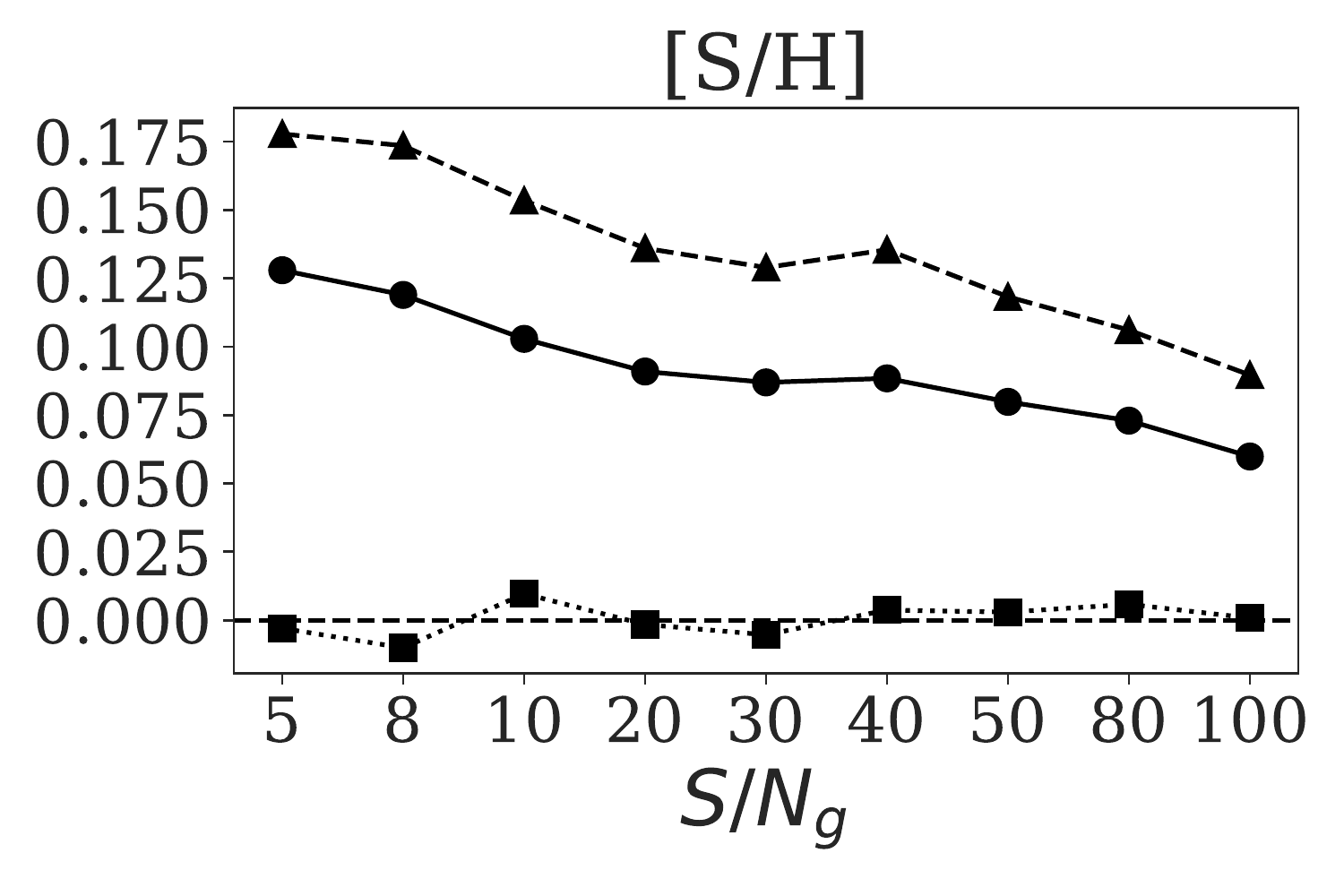}
  \includegraphics[width=0.23\textwidth]{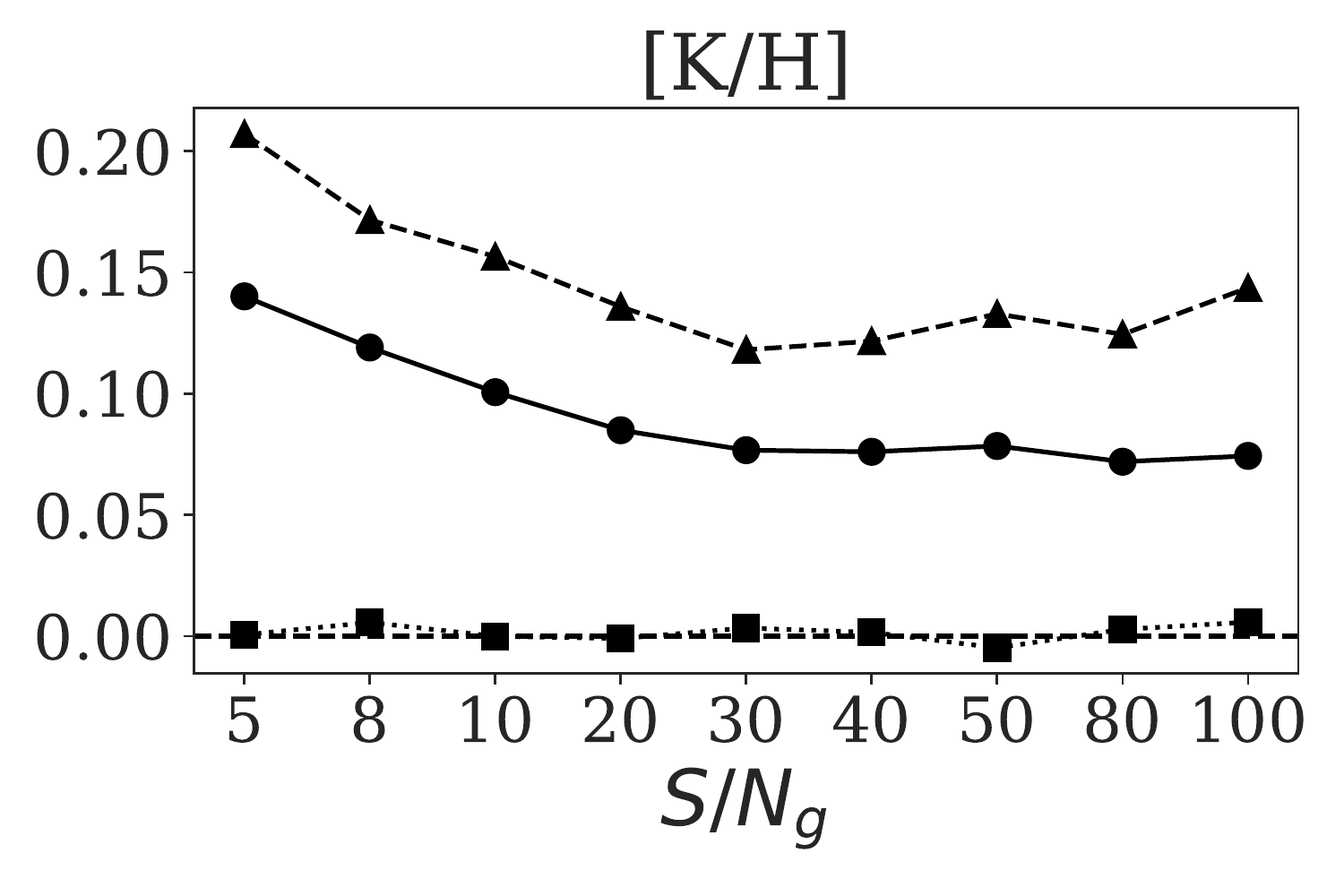}
  \includegraphics[width=0.23\textwidth]{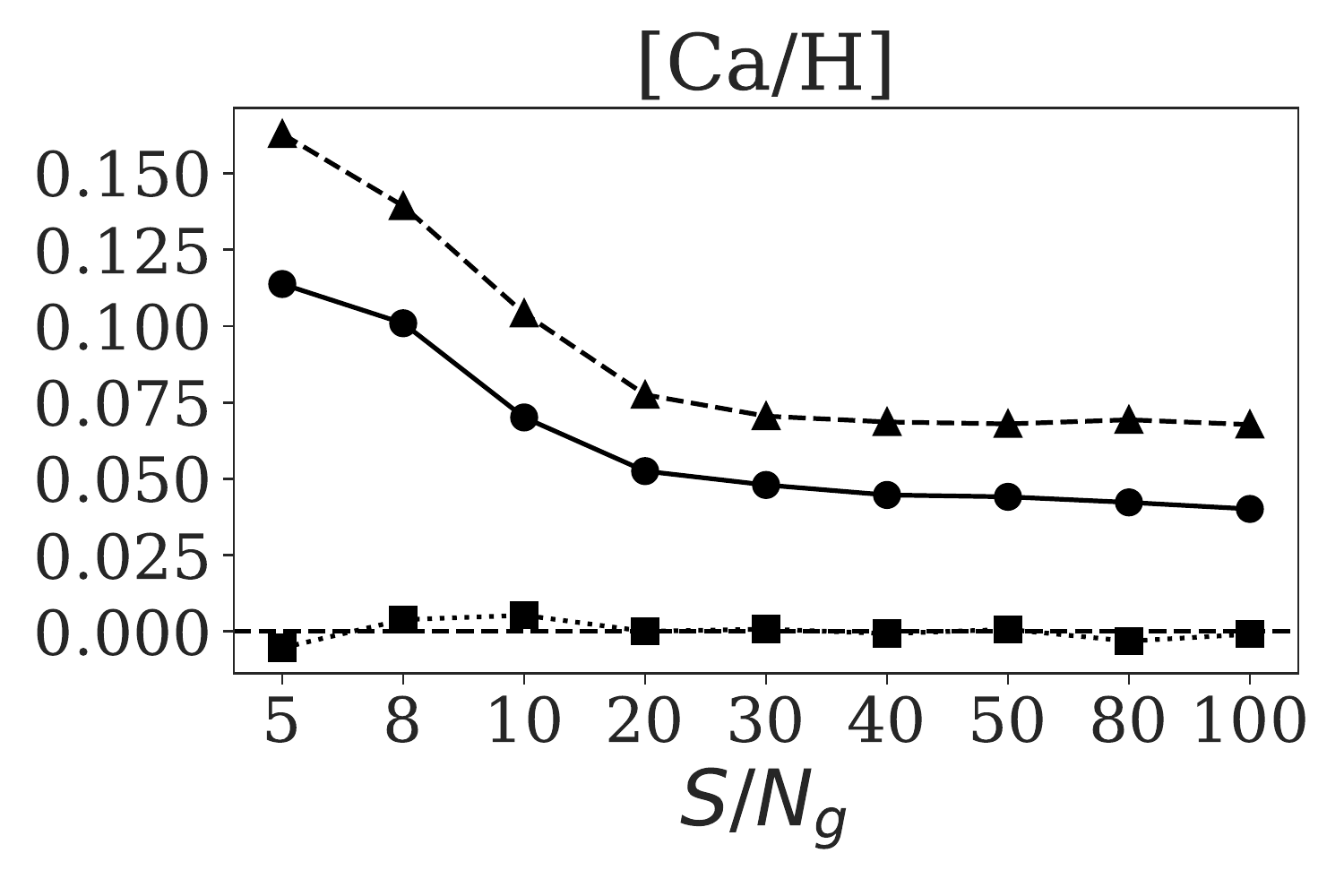}
  \includegraphics[width=0.23\textwidth]{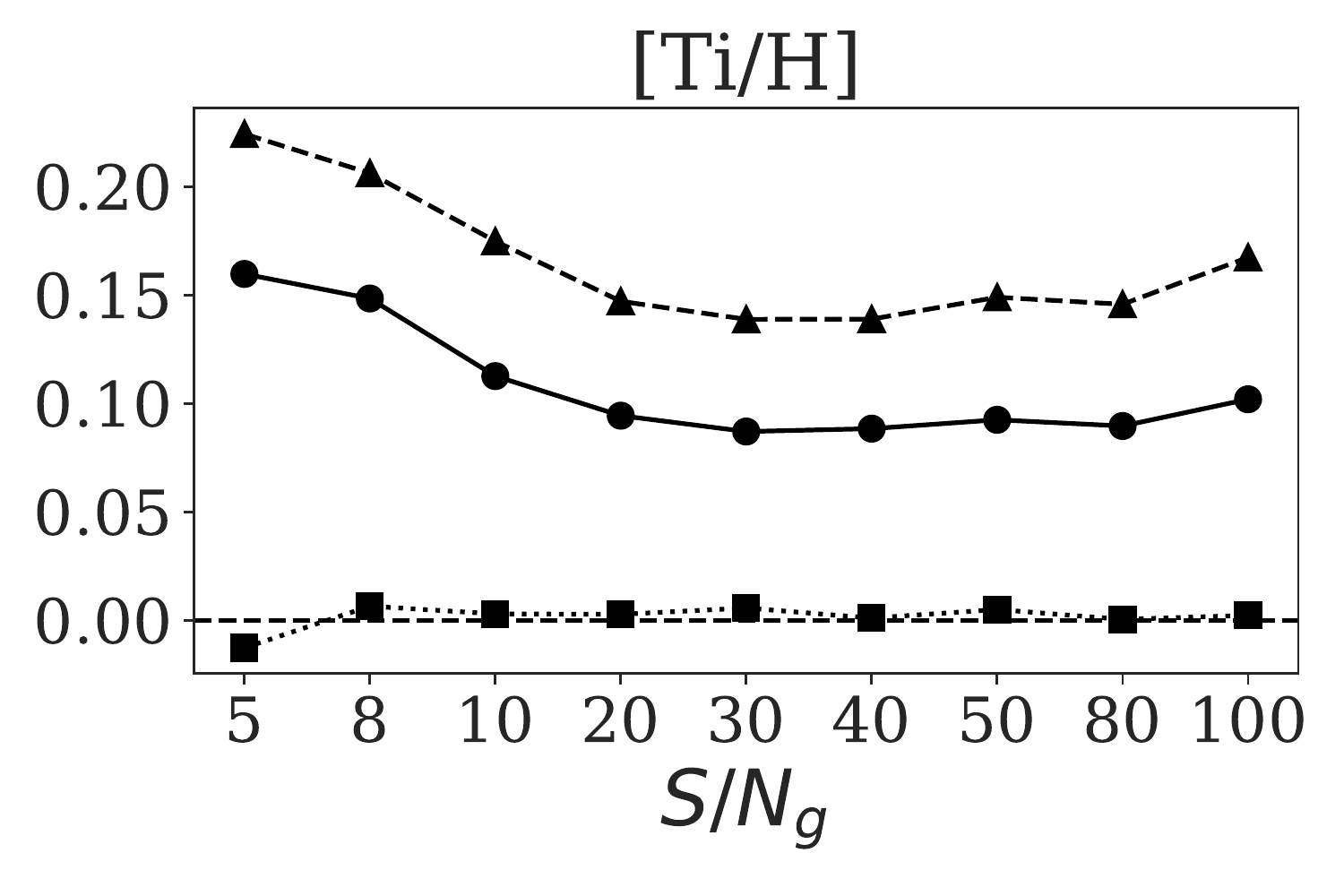}
  \includegraphics[width=0.23\textwidth]{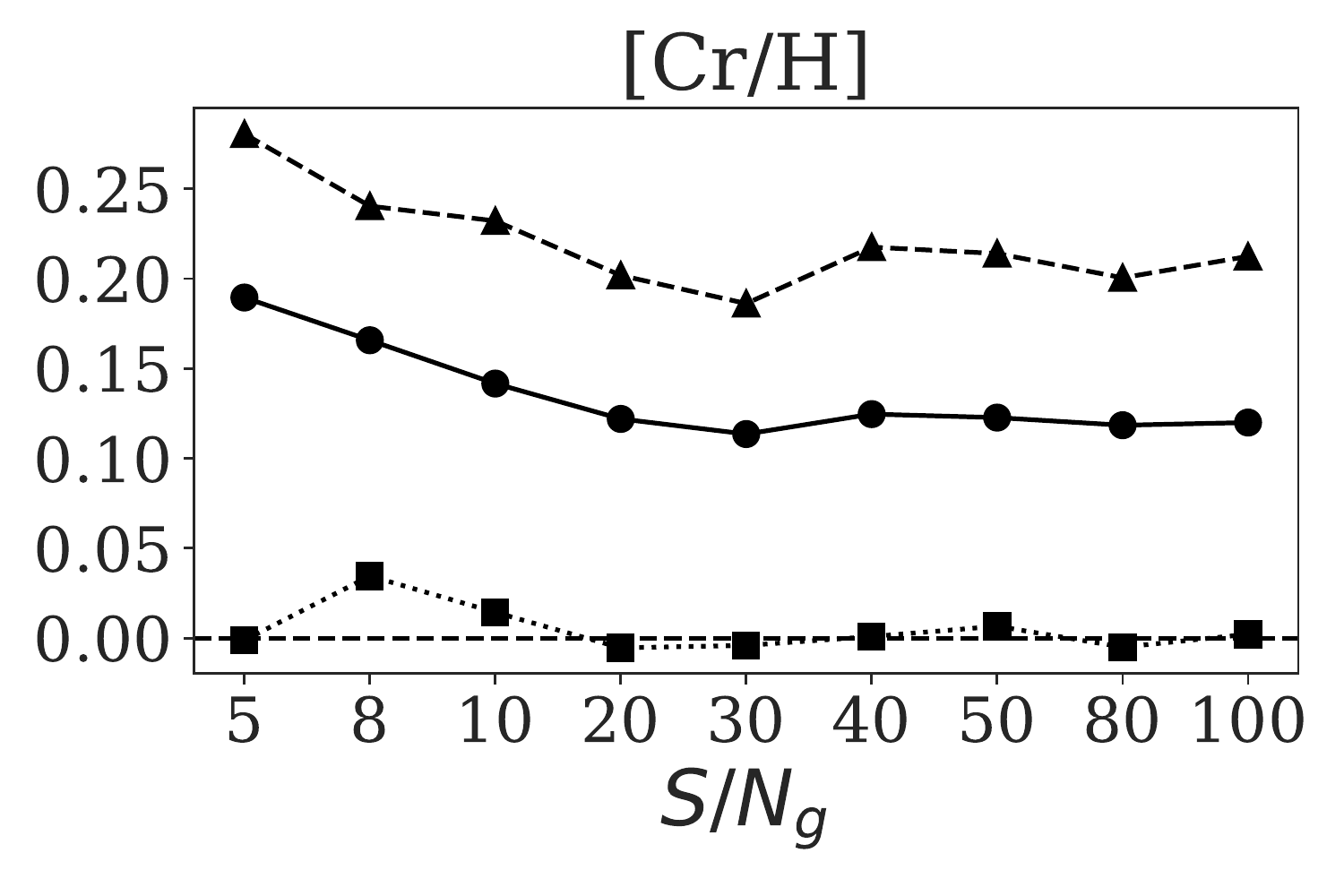}
  \includegraphics[width=0.23\textwidth]{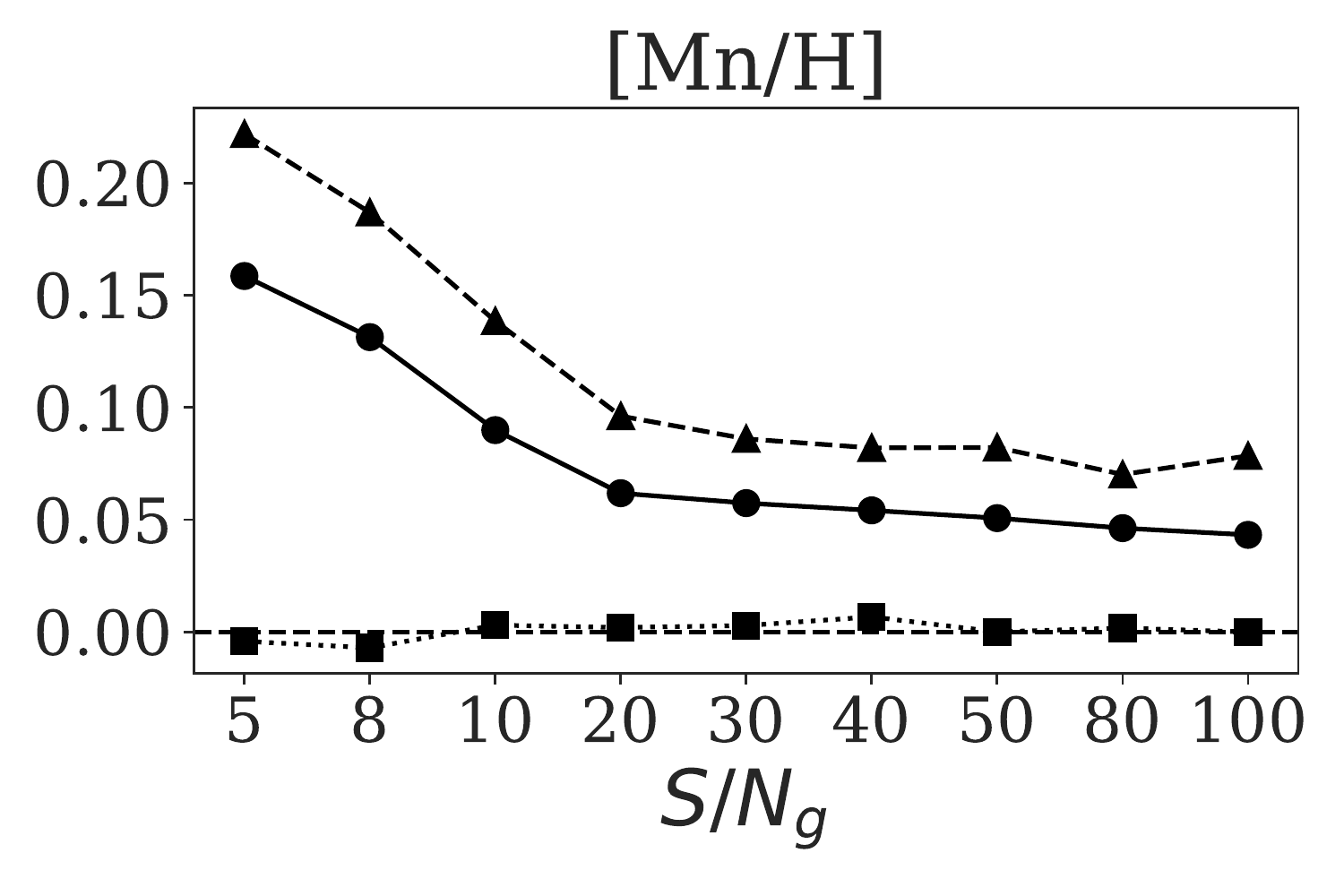}
  \includegraphics[width=0.23\textwidth]{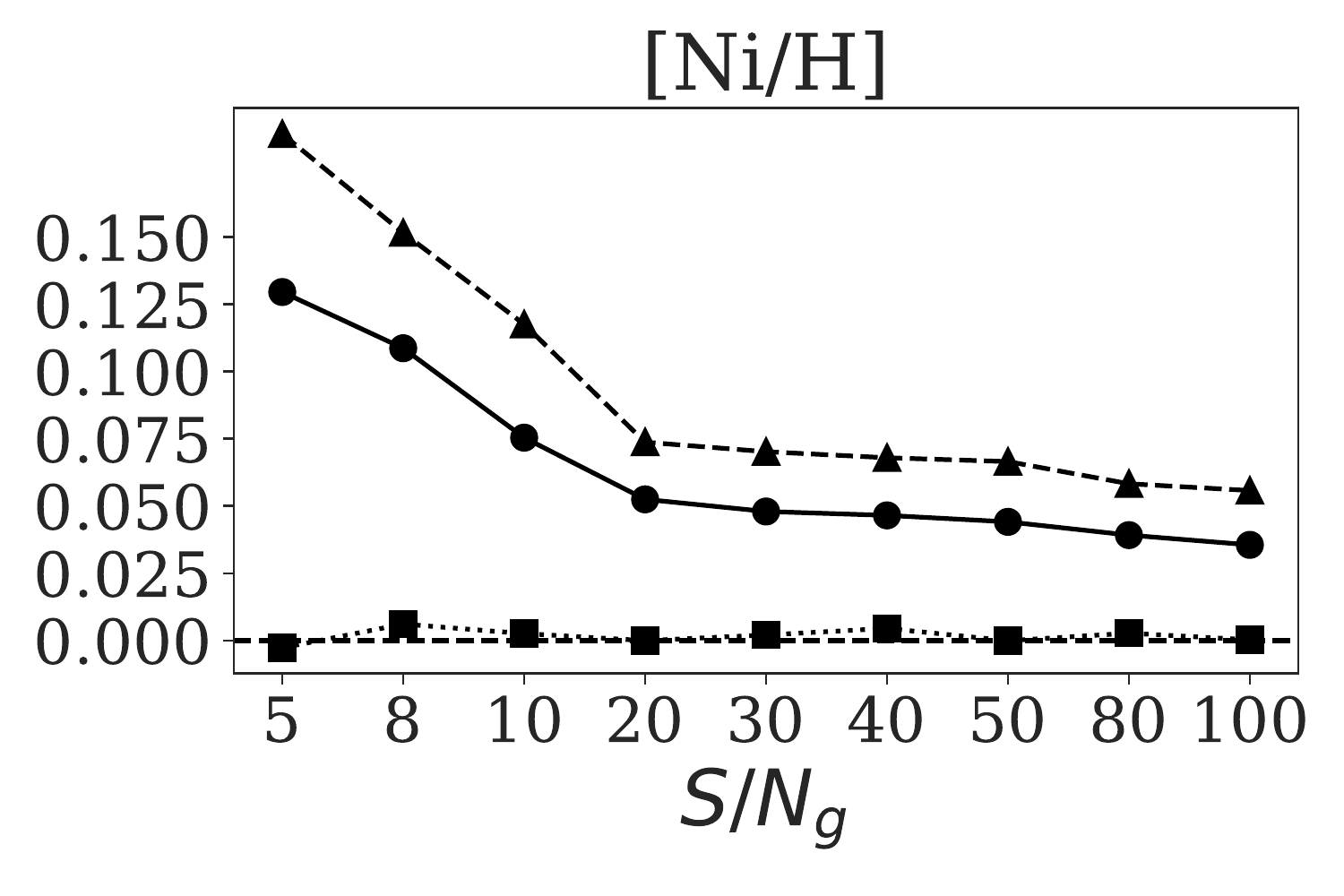}
  \caption{The dependence of the consistency between the StarGRUNet predictions and APOGEE DR17 catalog on the spectral signal-to-noise ratio. $\blacktriangle$, $\bullet$ and $\blacksquare$ represent $\sigma$, MAE and $\mu$ of the prediction uncertainty.}
  \label{fig111:widgets}
\end{figure*}

The prediction performance of StarGRUNet can also be measured by the dependence of the difference between its predictions and APOGEE DR17 catalog on the signal-to-noise ratio (Figure \ref{fig111:widgets}). The experiments in Figure \ref{fig111:widgets} investigate the dependence of the prediction error of StarGRUNet on the signal-to-noise ratio for the abundance of $13$ elements. The results show that the increase of $S/N_g$ can effectively reduce the MAE and $\sigma$ of StarGRUNet prediction, but $\mu$ is almost always stable at 0. These phenomena indicate that improving data quality can effectively reduce the error of StarGRUNet without affecting the overall consistency between StarGRUNet and APOGEE DR17 catalog. Therefore, the prediction results of StarGRUNet are very robust. In conclusion, the results of Figure \ref{fig8:widgets} and \ref{fig111:widgets} demonstrate the excellent prediction performance of StarGRUNet for all stellar parameters.

Finally, we compared the prediction results of StarGRUNet, StarNet, and ResNet \citep{2016_ResNet}. StarNet consists of several convolutional layers and several fully connected layers. Therefore, it is a typical convolutional neural network for stellar spectral parameter estimation. ResNet can mine the deep, longitudinal features of the spectrum. This is a sharp contrast to the cross-wavelength subbands feature extraction capability of StarGRUNet. Therefore, it helps us observe the advantages of cross-wavelength subband information extraction and fusion to compare these two methods with StarGRUNet. Table \ref{tab2:widgets} presents the experimental results of these three models. To fairly compare the experimental results of them, the three methods share the training set, validation set, and test set. The experimental results in Table \ref{tab2:widgets} indicate that StarGRUNet has an undeniable advantage. As for StarNet, its performance is much inferior to StarGRUNet. This is because the network structure of StarNet is relatively simple. Therefore, it is difficult to achieve better performance of parameter estimation on LAMOST low-resolution spectra. As for ResNet, although it can longitudinally exploit the weak features of the spectrum, it fails to extract the cross-band information horizontally, and the result of the failure is the ResNet's insufficiency on resistance to noise. In contrast, the prediction results of StarGRUNet are more accurate and robust. These results are not only due to the extraction and fusion of various cross-band feature information by the BGANet model but also due to the advantageous integration of the mapping results under multiple feature expressions by StarGRUNet.

\begin{table*}
\centering

\caption{\label{tab2:widgets} Comparisons: StarGRUNet, StarNet, and ResNet.}

\begin{tabular}{|c|ccc|ccc|ccc|}
\hline
\textbf{Model} &
  \multicolumn{3}{c|}{StarGRUNet} &
  \multicolumn{3}{c|}{StarNet} &
  \multicolumn{3}{c|}{ResNet} \\ \hline
\textbf{Error} &
  \multicolumn{1}{c|}{$\mu$} &
  \multicolumn{1}{c|}{$\sigma$} &
  MAE &
  \multicolumn{1}{c|}{$\mu$} &
  \multicolumn{1}{c|}{$\sigma$} &
  MAE &
  \multicolumn{1}{c|}{$\mu$} &
  \multicolumn{1}{c|}{$\sigma$} &
  MAE \\ \hline
$T_\texttt{eff}$(K) &
  \multicolumn{1}{c|}{0.935} &
  \multicolumn{1}{c|}{93.77} &
  49.28 &
  \multicolumn{1}{c|}{-33.41} &
  \multicolumn{1}{c|}{518.66} &
  416.97 &
  \multicolumn{1}{c|}{-16.80} &
  \multicolumn{1}{c|}{150.04} &
  87.69 \\ \hline
$\log \ g$(dex) &
  \multicolumn{1}{c|}{0.000} &
  \multicolumn{1}{c|}{0.162} &
  0.084 &
  \multicolumn{1}{c|}{-0.014} &
  \multicolumn{1}{c|}{0.984} &
  0.877 &
  \multicolumn{1}{c|}{0.023} &
  \multicolumn{1}{c|}{0.250} &
  0.146 \\ \hline
[Fe/H](dex) &
  \multicolumn{1}{c|}{0.001} &
  \multicolumn{1}{c|}{0.070} &
  0.041 &
  \multicolumn{1}{c|}{-0.003} &
  \multicolumn{1}{c|}{0.294} &
  0.224 &
  \multicolumn{1}{c|}{-0.008} &
  \multicolumn{1}{c|}{0.098} &
  0.058 \\ \hline
[C/H](dex) &
  \multicolumn{1}{c|}{0.001} &
  \multicolumn{1}{c|}{0.090} &
  0.055 &
  \multicolumn{1}{c|}{-0.002} &
  \multicolumn{1}{c|}{0.312} &
  0.218 &
  \multicolumn{1}{c|}{0.003} &
  \multicolumn{1}{c|}{0.115} &
  0.070 \\ \hline
[N/H](dex) &
  \multicolumn{1}{c|}{0.000} &
  \multicolumn{1}{c|}{0.182} &
  0.109 &
  \multicolumn{1}{c|}{-0.002} &
  \multicolumn{1}{c|}{0.375} &
  0.286 &
  \multicolumn{1}{c|}{0.008} &
  \multicolumn{1}{c|}{0.200} &
  0.122 \\ \hline
[O/H](dex) &
  \multicolumn{1}{c|}{0.000} &
  \multicolumn{1}{c|}{0.104} &
  0.068 &
  \multicolumn{1}{c|}{-0.002} &
  \multicolumn{1}{c|}{0.239} &
  0.176 &
  \multicolumn{1}{c|}{-0.002} &
  \multicolumn{1}{c|}{0.116} &
  0.075 \\ \hline
[Mg/H](dex) &
  \multicolumn{1}{c|}{0.001} &
  \multicolumn{1}{c|}{0.073} &
  0.045 &
  \multicolumn{1}{c|}{-0.001} &
  \multicolumn{1}{c|}{0.246} &
  0.178 &
  \multicolumn{1}{c|}{-0.004} &
  \multicolumn{1}{c|}{0.094} &
  0.059 \\ \hline
[Al/H](dex) &
  \multicolumn{1}{c|}{0.001} &
  \multicolumn{1}{c|}{0.089} &
  0.052 &
  \multicolumn{1}{c|}{-0.002} &
  \multicolumn{1}{c|}{0.311} &
  0.213 &
  \multicolumn{1}{c|}{0.009} &
  \multicolumn{1}{c|}{0.131} &
  0.079 \\ \hline
[Si/H](dex) &
  \multicolumn{1}{c|}{0.001} &
  \multicolumn{1}{c|}{0.074} &
  0.045 &
  \multicolumn{1}{c|}{-0.002} &
  \multicolumn{1}{c|}{0.258} &
  0.192 &
  \multicolumn{1}{c|}{-0.004} &
  \multicolumn{1}{c|}{0.095} &
  0.059 \\ \hline
[S/H](dex) &
  \multicolumn{1}{c|}{0.002} &
  \multicolumn{1}{c|}{0.121} &
  0.080 &
  \multicolumn{1}{c|}{-0.001} &
  \multicolumn{1}{c|}{0.236} &
  0.174 &
  \multicolumn{1}{c|}{0.005} &
  \multicolumn{1}{c|}{0.139} &
  0.088 \\ \hline
[K/H](dex) &
  \multicolumn{1}{c|}{0.002} &
  \multicolumn{1}{c|}{0.141} &
  0.082 &
  \multicolumn{1}{c|}{0.000} &
  \multicolumn{1}{c|}{0.276} &
  0.194 &
  \multicolumn{1}{c|}{-0.031} &
  \multicolumn{1}{c|}{0.236} &
  0.115 \\ \hline
[Ca/H](dex) &
  \multicolumn{1}{c|}{0.000} &
  \multicolumn{1}{c|}{0.081} &
  0.050 &
  \multicolumn{1}{c|}{-0.002} &
  \multicolumn{1}{c|}{0.255} &
  0.191 &
  \multicolumn{1}{c|}{0.003} &
  \multicolumn{1}{c|}{0.094} &
  0.058 \\ \hline
[Ti/H](dex) &
  \multicolumn{1}{c|}{0.002} &
  \multicolumn{1}{c|}{0.161} &
  0.101 &
  \multicolumn{1}{c|}{-0.001} &
  \multicolumn{1}{c|}{0.329} &
  0.246 &
  \multicolumn{1}{c|}{0.028} &
  \multicolumn{1}{c|}{0.226} &
  0.125 \\ \hline
[Cr/H](dex) &
  \multicolumn{1}{c|}{0.003} &
  \multicolumn{1}{c|}{0.215} &
  0.126 &
  \multicolumn{1}{c|}{-0.002} &
  \multicolumn{1}{c|}{0.384} &
  0.281 &
  \multicolumn{1}{c|}{0.013} &
  \multicolumn{1}{c|}{0.226} &
  0.134 \\ \hline
[Mn/H](dex) &
  \multicolumn{1}{c|}{0.001} &
  \multicolumn{1}{c|}{0.101} &
  0.060 &
  \multicolumn{1}{c|}{-0.003} &
  \multicolumn{1}{c|}{0.372} &
  0.280 &
  \multicolumn{1}{c|}{-0.006} &
  \multicolumn{1}{c|}{0.126} &
  0.077 \\ \hline
[Ni/H](dex) &
  \multicolumn{1}{c|}{0.000} &
  \multicolumn{1}{c|}{0.082} &
  0.050 &
  \multicolumn{1}{c|}{0.000} &
  \multicolumn{1}{c|}{0.326} &
  0.241 &
  \multicolumn{1}{c|}{0.002} &
  \multicolumn{1}{c|}{0.108} &
  0.064 \\ \hline
\end{tabular}
\end{table*}

\subsection{Model Uncertainty}\label{Sec:Methods:ModelUncertainty}

Model uncertainty analysis is another way to test the prediction performance of StarGRUNet. {\citet{2016_Dropout}} demonstrated that a neural network with a Dropout mechanism is an approximation to a Bayesian neural network with a Gaussian distribution and can be employed to estimate prediction uncertainty. \citet{2018_AstroNN} introduced this idea into a stellar spectral parameters estimation model, AstroNN, to estimate the uncertainty. Since BGANet comes with a Dropout mechanism in each of the hidden feature vectors, it supports our assessment of the model uncertainty. We repeatedly estimated each spectral parameter for five times from each stellar spectrum, and computed the standard deviation of the five prediction as the model uncertainty of StarGRUNet.

\begin{figure*}
  \centering
  \includegraphics[width=0.8\textwidth]{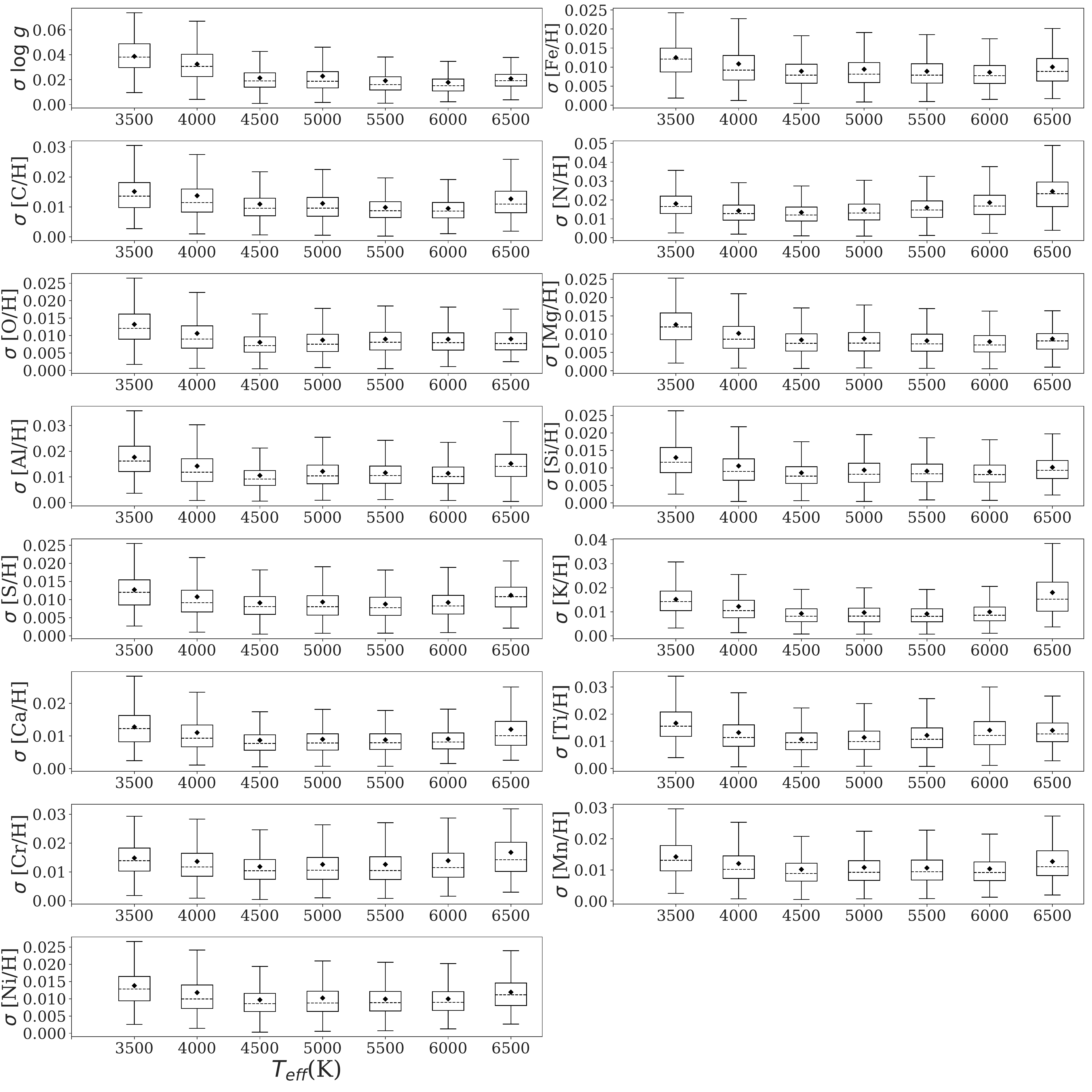}
  \caption{ The dependencies of StarGRUNet prediction uncertainty on $T_\texttt{eff}$. The dependencies are presented using box plots. The black dots inside the box represent the mean of prediction uncertainties. The dashed line inside the box represents the second quartile of the prediction uncertainty $Q_2$ (namely the median). The lower bottom of the box represents the first quartile $Q_1$. The upper bottom of the box represents the third quartile $Q_3$. The difference between $Q_3$ and $Q_1$ is called $IQR$ (interquartile range): $IQR=Q_3-Q_1$. The lines above and below the box are called the upper and lower limits, corresponding to the values $Q_3+1.5IQR$ and $Q_1-1.5IQR$, respectively. The height of the box and the distance between the upper limit and the lower limit can reflect the degree of uncertainty dispersion to some extent.}
  \label{fig15:widgets}
\end{figure*}

\begin{figure*}
  \centering
  \includegraphics[width=0.7\textwidth]{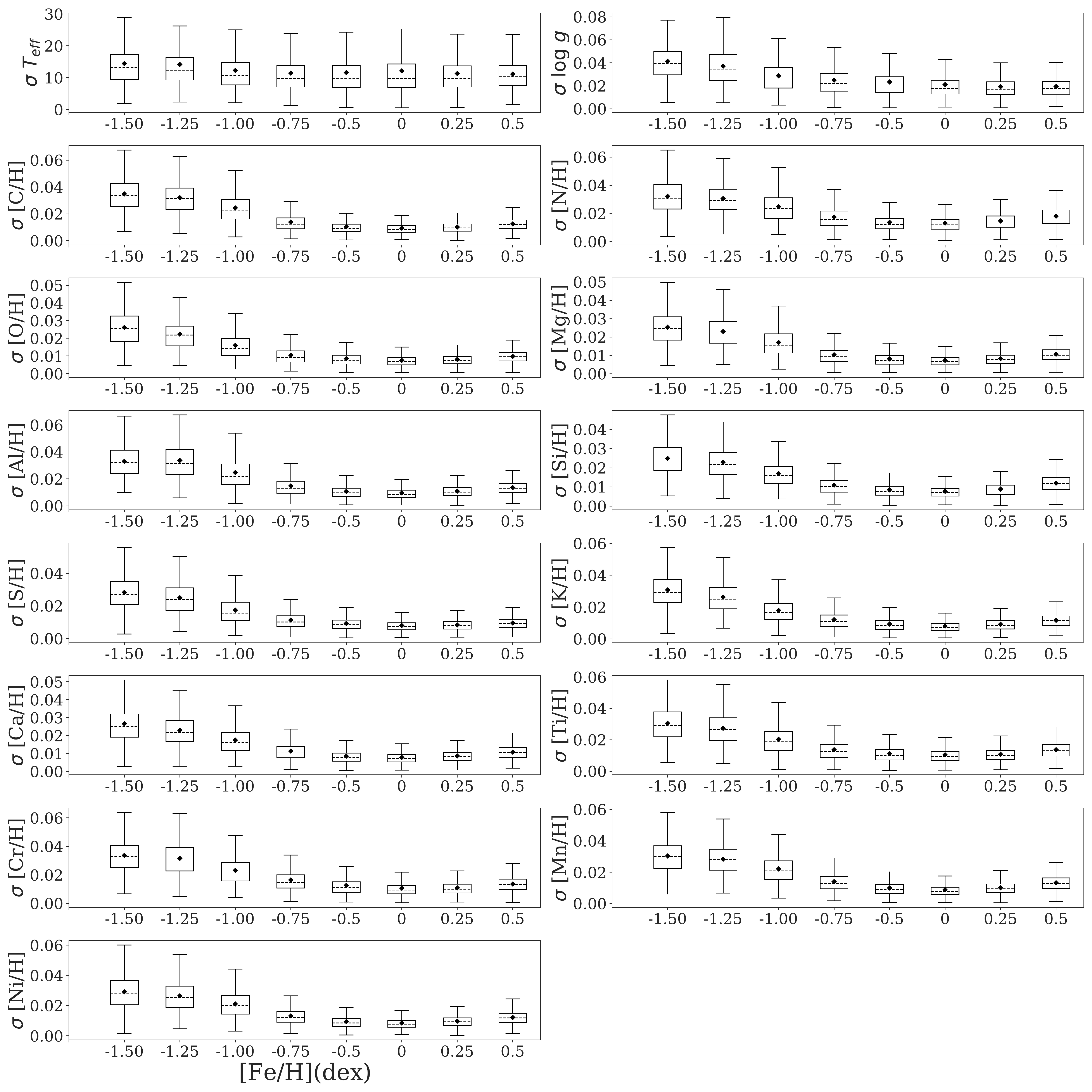}
  \caption{The dependencies of StarGRUNet prediction uncertainty on [Fe/H]. The dependencies are presented using box plots. The black dots inside the box represent the mean of prediction uncertainties. The dashed line inside the box represents the second quartile of the prediction uncertainty $Q_2$ (namely the median). The lower bottom of the box represents the first quartile $Q_1$. The upper bottom of the box represents the third quartile $Q_3$. The difference between $Q_3$ and $Q_1$ is called $IQR$ (interquartile range): $IQR=Q_3-Q_1$. The lines above and below the box are called the upper and lower limits, corresponding to the values $Q_3+1.5IQR$ and $Q_1-1.5IQR$, respectively. The height of the box and the distance between the upper limit and the lower limit can reflect the degree of uncertainty dispersion to some extent.
  }
  \label{fig16:widgets}
\end{figure*}

Figures \ref{fig15:widgets} and \ref{fig16:widgets} present the dependence of StarGRUNet uncertainty on $T_\texttt{eff}$ and [Fe/H], respectively. It is shown that the parameter estimation uncertainty of StarGRUNet on the spectra of metal-poor star ([Fe/H]<-1.00dex), cold star ($T_\texttt{eff}<4000K$) and hot star ($T_\texttt{eff}> 6000$K) is generally greater than that on other types of spectra. These trends are in general consistent with \citet{2018_AstroNN}. The reason for this phenomenon is the small number and weak spectral features of metal-poor, cool, and hot stars. These characteristics reduce the performance of the learned model on spectra of these stars. Therefore, we recommend using the results in these above-mentioned ranges with caution.

\section{Applications on LAMOST DR8 Low-resolution Spectra and Validations on Other Surveys }\label{Sec:Applications}

\subsection{Applications on LAMOST DR8 Low-resolution Spectra}

In Section \ref{Sec:Methods}, we performed a comprehensive evaluation on the performance of StarGRUNet and a series of experimental results indicate its effectiveness and robustness. Therefore, we utilized the trained StarGRUNet models to estimate the stellar parameters $T_\texttt{eff}$ and $\log~g$, $14$ chemical elemental abundances, and $1\sigma$ uncertainties for about 8.21 million LAMOST low-resolution spectra with $S/N_g\ge5$, and generated the StarGRUNet-LAMOST catalog. In subsections \ref{Sec:Applications:Consistency}, \ref{Sec:Applications:CatalogsComparisons} and \ref{Sec:Applications:Uncertainties}, we evaluated the reliability of StarGRUNet-LAMOST catalog.

\subsection{Consistencies with GALAH Survey}\label{Sec:Applications:Consistency}

\begin{figure*}
  \centering
  \includegraphics[width=0.3\textwidth]{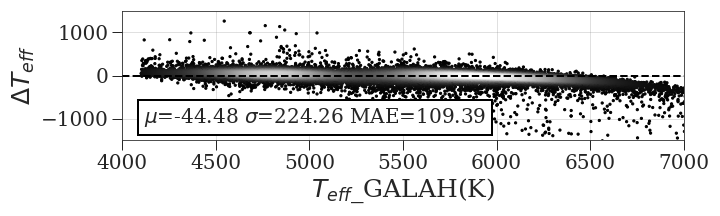}
  \includegraphics[width=0.3\textwidth]{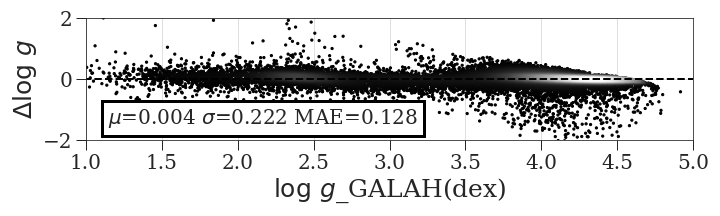}
  \includegraphics[width=0.3\textwidth]{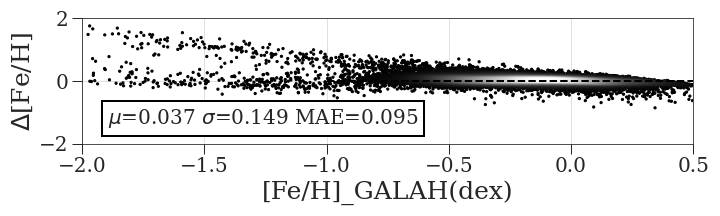}
  \includegraphics[width=0.3\textwidth]{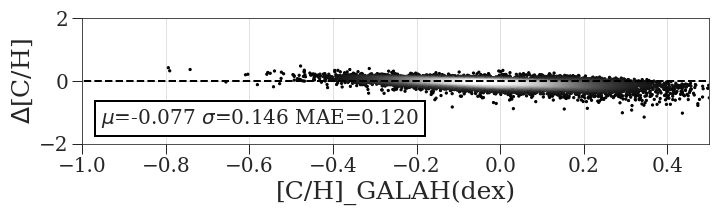}
  \includegraphics[width=0.3\textwidth]{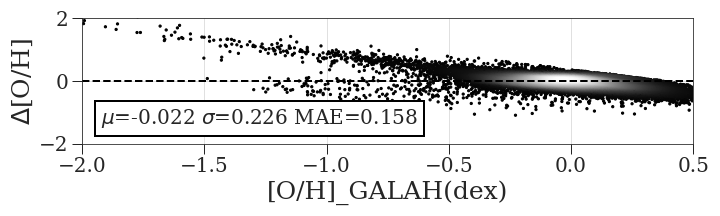}
  \includegraphics[width=0.3\textwidth]{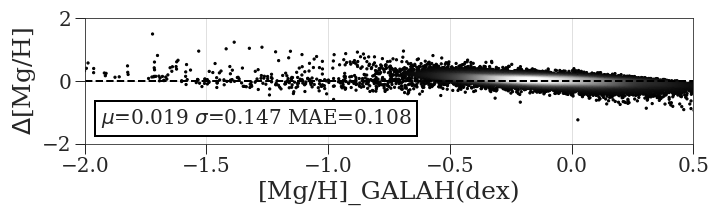}
  \includegraphics[width=0.3\textwidth]{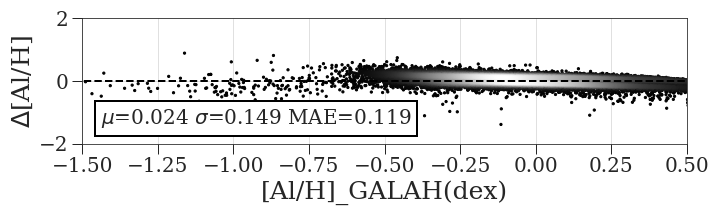}
  \includegraphics[width=0.3\textwidth]{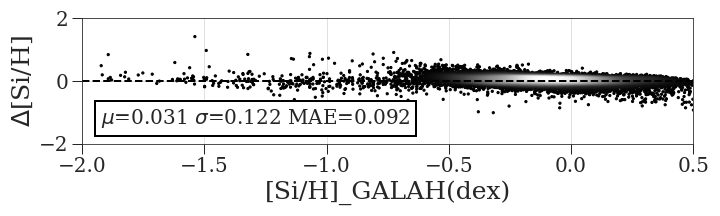}
  \includegraphics[width=0.3\textwidth]{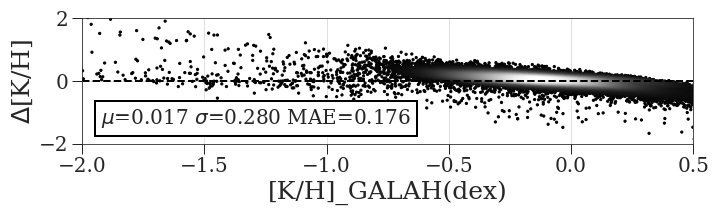}
  \includegraphics[width=0.3\textwidth]{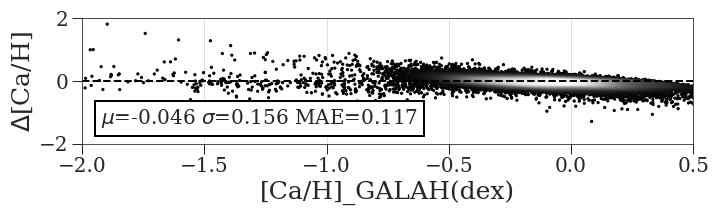}
  \includegraphics[width=0.3\textwidth]{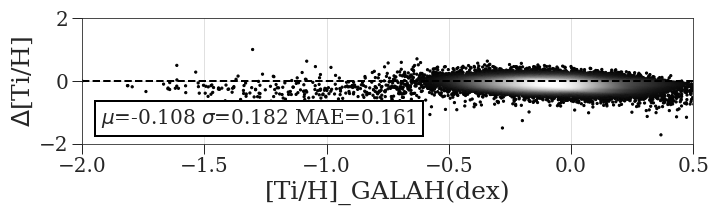}
  \includegraphics[width=0.3\textwidth]{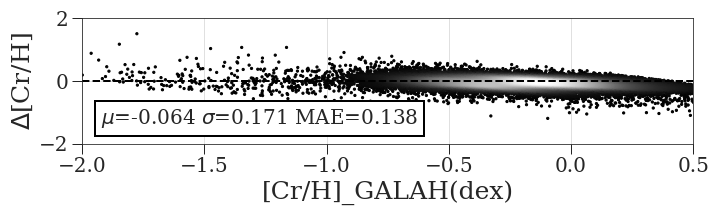}
  \includegraphics[width=0.3\textwidth]{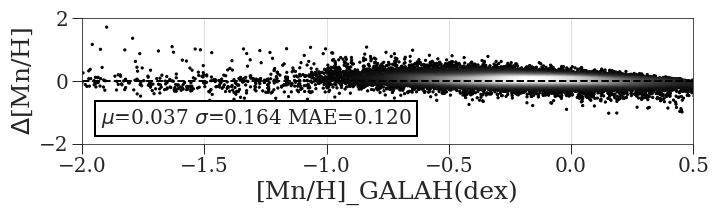}
  \includegraphics[width=0.3\textwidth]{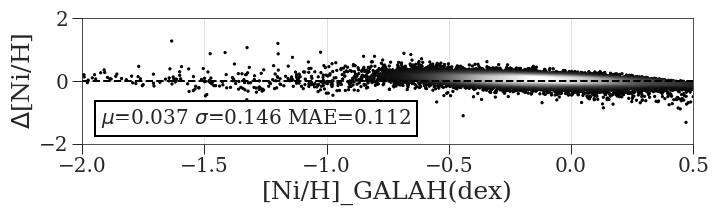}
  \caption{Consistency between StarGRUNet-LAMOST catalog and GALAH DR3 catalog. In each subplot, the horizontal axis indicates the results provided by GALAH DR3 catalog, and the vertical axis indicates the difference between the StarGRUNet-LAMOST catalog and the GALAH DR3 catalog. The dashed line corresponds to $\mu=0$, indicating the theoretical consistency. The lower left corner labels the bias and dispersion. The color characterizes the density of the samples.}
  \label{fig12:widgets}
\end{figure*}

It is an effective way to verify the reliability of the computed StarGRUNet-LAMOST catalog to investigate its consistency one one high-resolution catalog. GALAH DR3 \citep{2021_galah_dr3} provides reliable stellar parameters and elemental abundances for 588,571 stars, including 383,088 dwarfs, 200,927 giant stars, and 4,556 unclassified stars. We cross-matched the GALAH DR3 catalog with the StarGRUNet-LAMOST catalog and obtained 27,527 common sources. Based on these common sources, we computed the consistency between the StarGRUNet-LAMOST catalog and the GALAH DR3 catalog (Figure \ref{fig12:widgets}). It is shown that there exists a high consistency between the StarGRUNet-LAMOST catalog and the GALAH DR3 catalog. The systematic biases on $T_\texttt{eff}$, $\log~g$ and [Fe/H] are $-44.48$K, $0.004$dex, and $0.037$dex, respectively. The corresponding dispersions are $224.26$ K, $0.222$ dex, and $0.149$ dex, respectively. The corresponding MAEs are $109.39$ K, $0.128$ dex, and $0.095$ dex, respectively.
The  biases, the deviations, and the MAEs of other elemental abundances are also similarly small. These experimental results show excellent consistency between the StarGRUNet-LAMOST catalog and the GALAH DR3 catalog, and indicate the reliability of the StarGRUNet-LAMOST catalog.

\begin{figure*}
  \centering
  \includegraphics[scale=0.256]{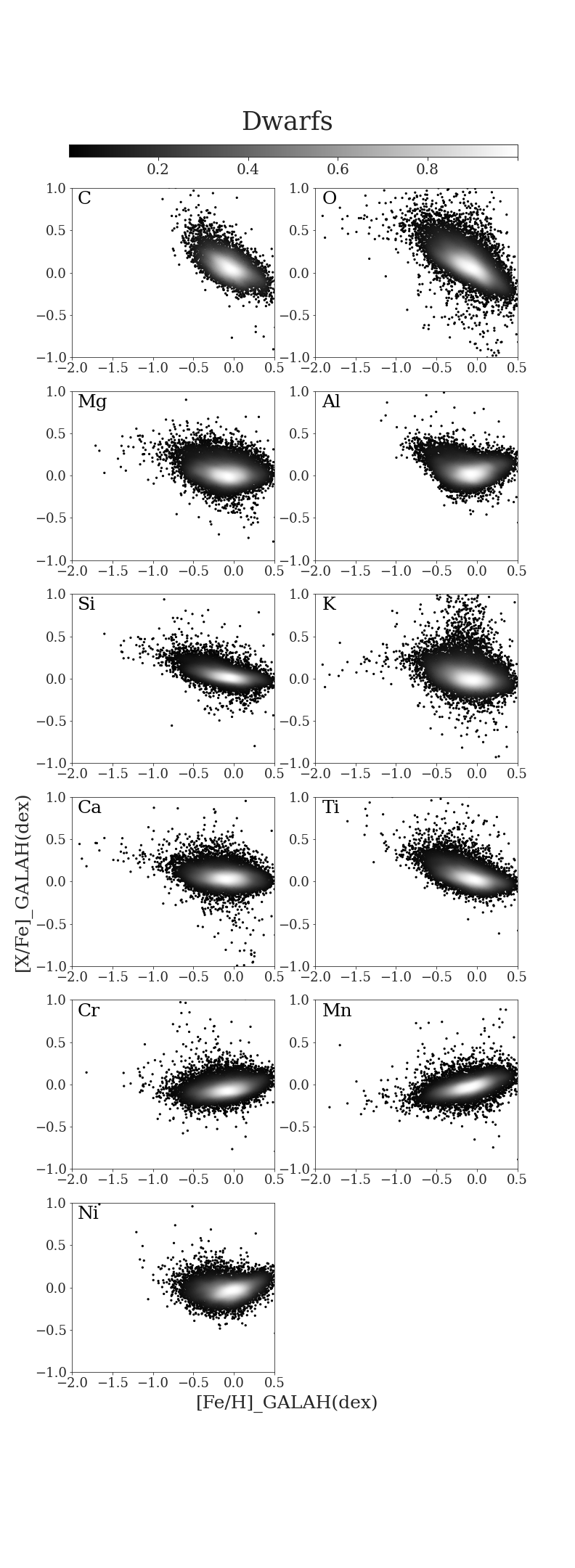}
  \includegraphics[scale=0.256]{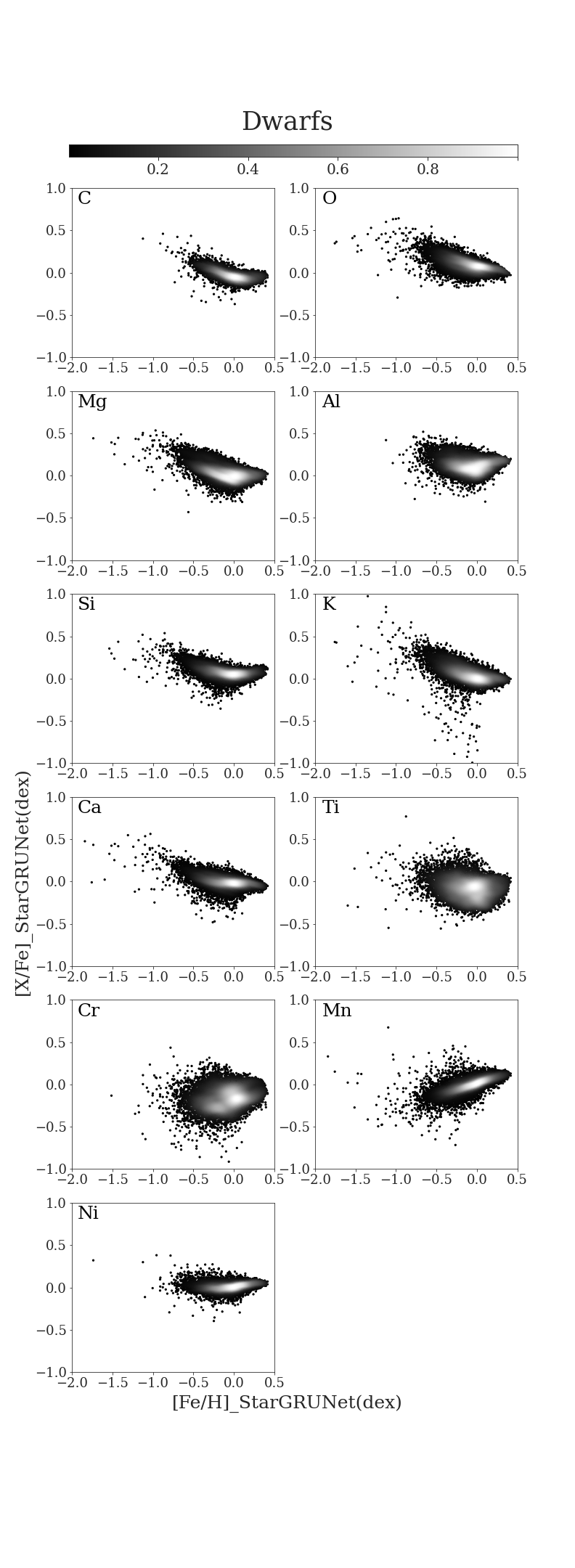}
   \vspace{-1.50cm}
  \caption{Distribution of dwarfs ($\log\ g>4$) in [X/Fe]-[Fe/H] space. The two left columns are the estimation results of the GALAH catalog, and the two right columns are the estimation results of the StarGRUNet-LAMOST catalog. The color characterizes the density of the sample distribution.}
  \label{fig13:widgets}
\end{figure*}

\begin{figure*}
  \centering
  \includegraphics[scale=0.256]{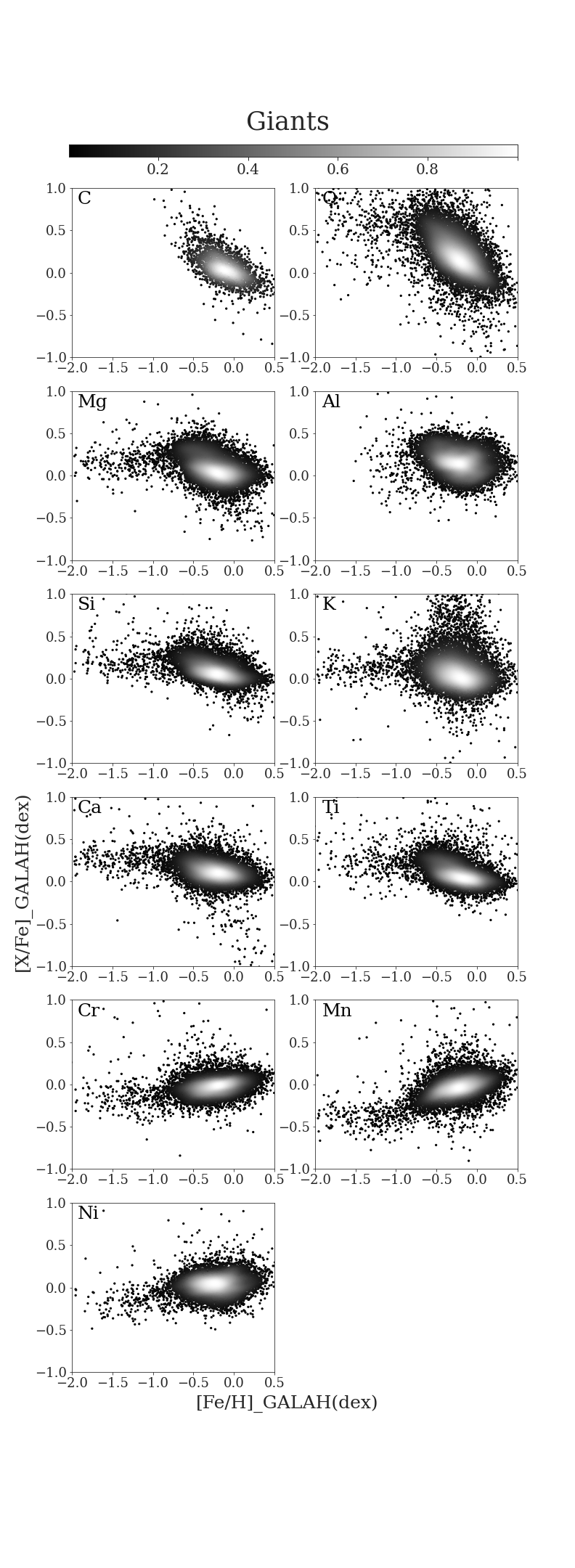}
  \includegraphics[scale=0.256]{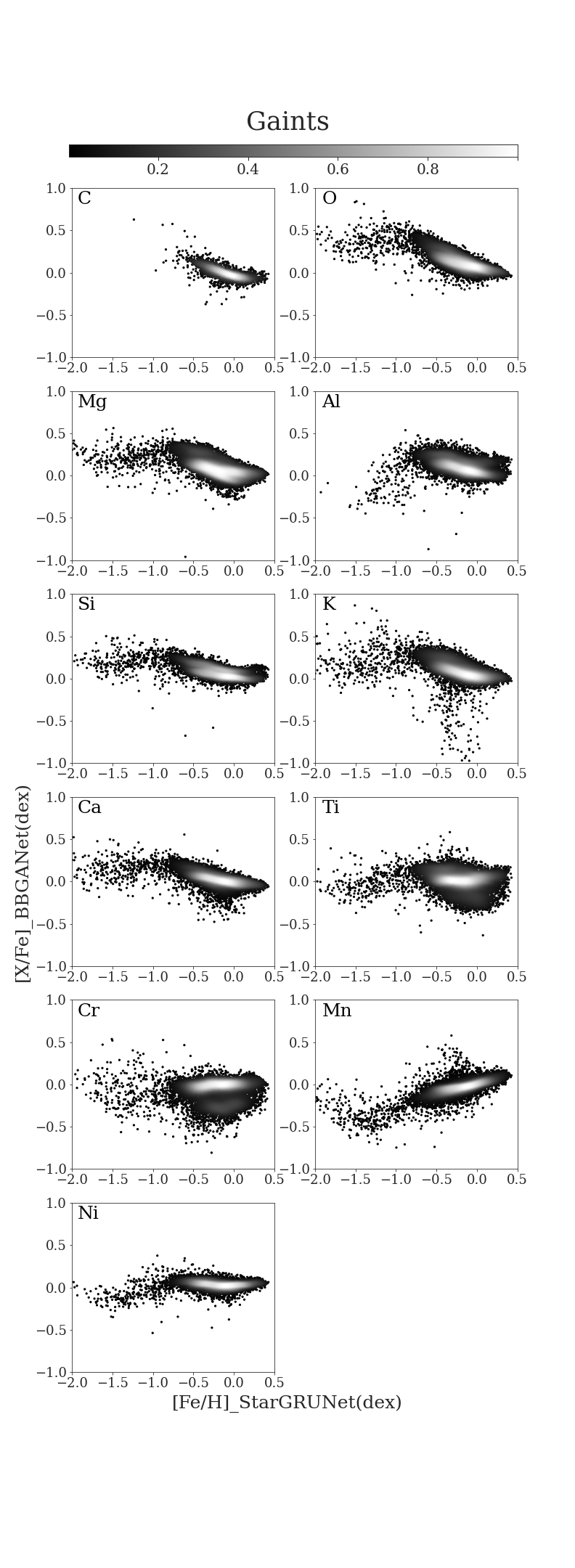}
   \vspace{-1.50cm}
  \caption{Distribution of giants ($\log\ g<4$) in the [X/Fe]-[Fe/H] space. The two left columns are the estimation results of the GALAH catalog, and the two right columns are the estimation results of the StarGRUNet-LAMOST catalog. The color characterizes the density of the sample distribution.}
  \label{fig14:widgets}
\end{figure*}

Figure \ref{fig13:widgets} and Figure \ref{fig14:widgets} show the distribution of Dwarf stars and Giant stars in [X/Fe]-[Fe/H] space, respectively. [X/Fe] represents the abundance of element X relative to Fe, and is computed as [X/Fe] = [X/H] - [Fe/H]. In general, the elemental abundances of the StarGRUNet-LAMOST catalog are relatively tight, and most of the StarGRUNet-LAMOST elemental abundances are consistent with the GALAH DR3 catalog. However, there are still some evident differences between the StarGRUNet-LAMOST catalog and the GALAH DR3 catalog on some elemental abundances, such as [Ti/H] of the Dwarfs. Such differences may be due to the severe lack of metal lines of these elements in the low-resolution, blue-end spectra of LAMOST.
Therefore, the precision of the Ti abundance of Dwarfs in the StarGRUNet-LAMOST catalog may be inferior to that in the GALAH DR3 catalog and should be used with caution.

\subsection{Comparisons with Other Catalogs based on LAMOST Low-resolution Spectra}\label{Sec:Applications:CatalogsComparisons}

To evaluate the effectiveness of the StarGRUNet-LAMOST catalog, this work compared it with three catalogs based on LAMOST low-resolution spectra. The three catalogs are the LASP catalog \citep{2015_LASP}, the GSN catalog \citep{2019_GSN}, and the LASSO-MLPNet catalog \citep{LiXiangruMNRAS2022}. The LASP catalog is computed by the LAMOST official pipeline and consists of the estimations of the stellar atmospheric parameters $T_\texttt{eff}$, $\log~g$, and [Fe/H]. The GSN catalog is a set of the estimates of the stellar atmospheric parameters $T_\texttt{eff}$, $\log~g$, [Fe/H], and [$\alpha$/Fe]. The LASSO-MLPNet catalog consists of the estimates of the stellar atmospheric parameters $T_\texttt{eff}$, $\log~g$, and [Fe/H] from $4,828,190$ LAMOST DR8 low-resolution stellar spectra with $5\le S/N_g\le 80$ and $3500$ K $\le T_\texttt{eff}\le$ $6500$ K. Since these catalogs are computed from LAMOST low-resolution spectra, they are very comparable with the StarGRUNet-LAMOST catalog.

\begin{figure*}
  \centering
  \includegraphics[width=0.9\textwidth]{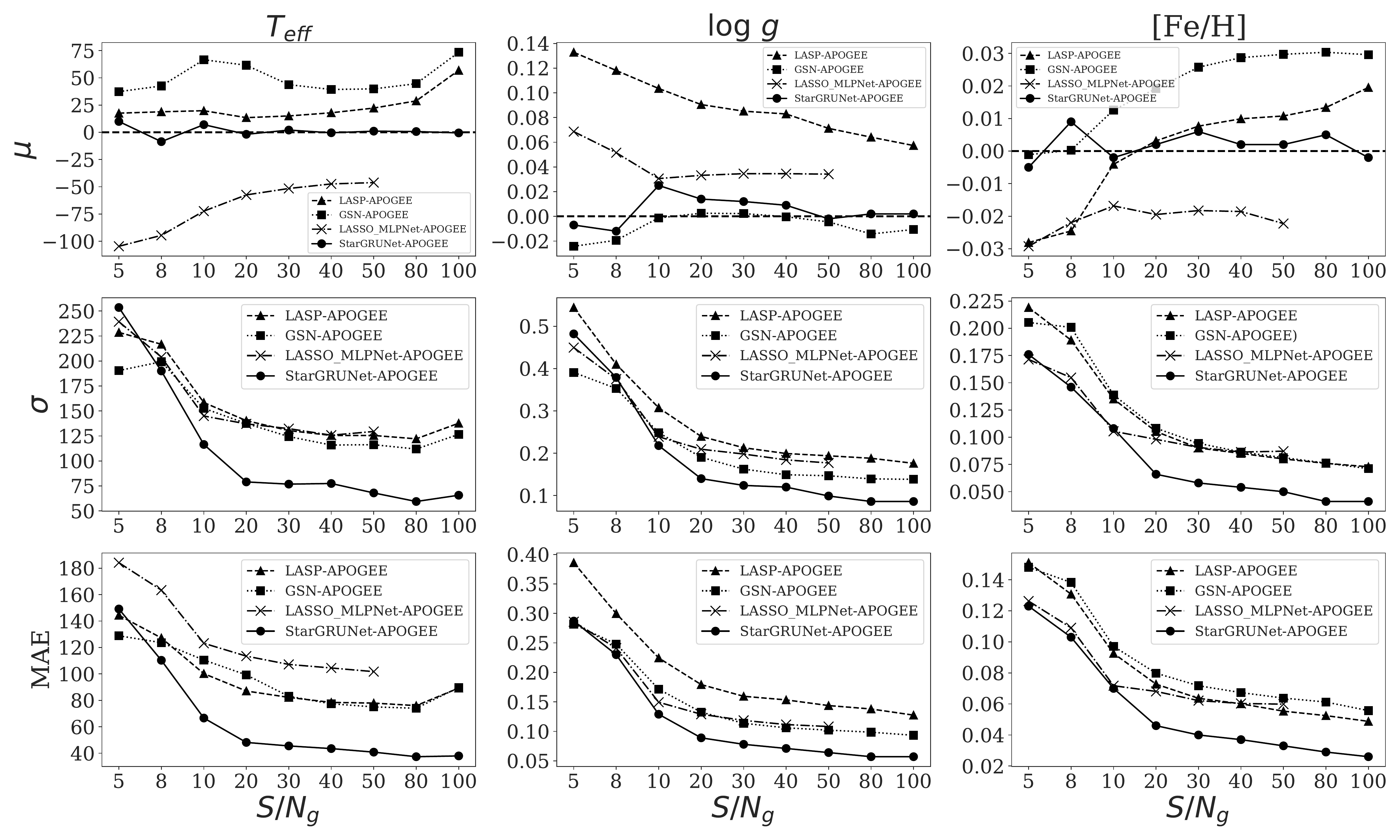}
  \caption{ Dependencies of the prediction errors on the spectral signal-to-noise ratio for the StarGRUNet-LAMOST catalog, the LASP catalog \citep{2015_LASP}, the GSN catalog \citep{2019_GSN}, and the LASSO-MLPNet catalog \citep{LiXiangruMNRAS2022}. The horizontal coordinates represent the $S/N_g$ intervals [5,8), [8,10), [10,20), [20,30), [30,40), [40,50), [50,80), [80,100), and [100,+$\infty$), respectively. The first, second and third columns represent ${T}_\texttt{eff}$, $\log\ g$ and [Fe/H], respectively. The first, second, and third rows respectively represent the mean  $\mu$, the standard deviation $\sigma$, and the mean of the absolute error MAE of the difference between (LASP catalog, GSN catalog, LASSO MLPNet catalog, StarGRUNet catalog) and APOGEE DR17 catalog. $\blacktriangle$, $\blacksquare$, $\times$and $\bullet$ indicate the evaluation results for the LASP catalog, the GSN catalog, the LASSO-MLPNet catalog, and the StarGRUNet-LAMOST catalog, respectively. It should be noted that the LASSO-MLPNet catalog only gives estimates for spectra with $5\le S/N_g\le80$. Therefore, the curves of the LASSO-MLPNet catalog disappear on the last two $S/N_g$ intervals.}
  \label{fig10:widgets}
\end{figure*}

In the experiment of Figure \ref{fig10:widgets}, we compared the StarGRUNet-LAMOST catalog with the LASP catalog, the GSN catalog \citep{2019_GSN}, and the LASSO-MLPNet catalog \citep{LiXiangruMNRAS2022} using the mean of the error, the standard deviation of the error, and MAE of the prediction error. It is shown that for each stellar atmosphere parameter, the estimation performance measures $|\mu|$, $\sigma$, and MAE of StarGRUNet are evidently lower than those of LASP, GSN, and LASSO-MLPNet on the whole. These results indicate that the error between the StarGRUNet-LAMOST catalog and the APOGEE DR17 catalog is smaller than that of other catalogs. Therefore, the StarGRUNet-LAMOST catalog can more accurately recover the stellar atmospheric parameters from LAMOST low-resolution spectra.

\subsection{Uncertainty Analysis Based on Repeated Observation: Observation Uncertainty}\label{Sec:Applications:Uncertainties}

We explored the model uncertainty of StarGRUNet based on the dropout technique in subsection \ref{Sec:Methods:ModelUncertainty}. In addition, LAMOST produced some repeated observations by carrying out multiple observations on some stars at various times and under different conditions. The parameters of this kind spectra from a common source can be assumed to be constant over the time span in which we carry out the observations. Therefore, such repeated observations provide us with an alternative option for analyzing the uncertainty of the StarGRUNet-LAMOST catalog. For convenience, we
name this uncertainty as observation uncertainty. Suppose the number of repeated observations of a star is $n_{obs}$, and the corresponding repeated spectra are \{$x_{1},x_{2},...,x_{n_{obs}}$\}. Thus, for anyone stellar parameter, StarGRUNet computed $n_{obs}$ estimates. The observation uncertainty is measured using the standard deviation of these $n_{obs}$ estimates in this work. To ensure the reliability of the estimated uncertainty, we only keep the target stars with more than six repeated observations ($26,459$ in total).

\begin{figure*}
  \centering
  \includegraphics[width=0.7\textwidth]{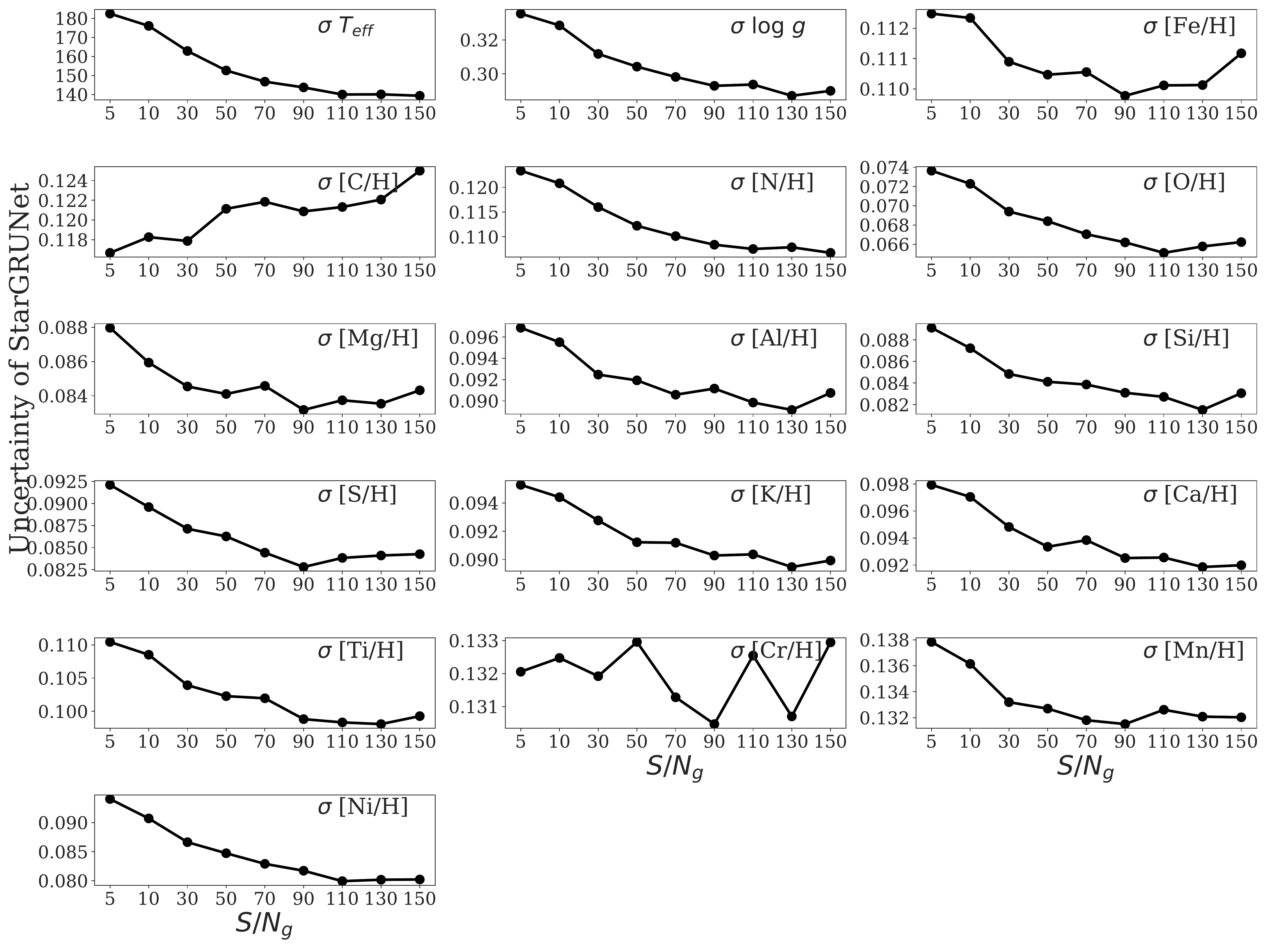}
  \caption{Observation uncertainty of StarGRUNet-LAMOST. The horizontal axis represents the signal-to-noise ratio of the spectra. Each subplot is labeled with the name of the corresponding stellar parameter or elemental abundance in the upper right corner.}
  \label{fig17:widgets}
\end{figure*}

Figure \ref{fig17:widgets} demonstrates the dependence of the observation uncertainty on the signal-to-noise ratio. Overall, the observation uncertainty of StarGRUNet-LAMOST catalog is low and has a clear decreasing trend with the increasing of spectrum quality. In the case of $S/N_g\in[5,10)$, the uncertainties of $T_\texttt{eff}$ and $\log \ g$ estimations are $182$ K and $0.34$ dex, respectively, and the uncertainty of elemental abundance estimation is $0.07$ dex-$0.17$ dex. In the case of $S/N_g\ge 150$, the uncertainty of $T_\texttt{eff}$ and $\log \ g$ estimates drop to $139$ K and $0.29$ dex, respectively, and the uncertainty of elemental abundance estimation drop to $0.06$ dex to $0.15$ dex. These phenomena indicate that the results of the StarGRUNet-LAMOST catalog are very robust.

\section{Conclusion}\label{Sec:Conclusion}

In this paper, a novel spectral parameter estimation neural network, BGANet, is designed based on Bi-GRU and Self-Attention mechanism. The parameter estimation performance is further improved by introducing an ensemble learning method StarGRUNet based on the BGANet. The competitiveness of the proposed method was evaluated by comparing it with the typical methods RNN, GRU, Bi-GRU, and StarNet.

By cross-matching the LAMOST DR8 low-resolution spectral library with the APOGEE DR17 catalog, we established a training set, a validation set, and a test set. These datasets are released for algorithm research, and used for learning and testing the proposed scheme. On the spectra with $S/N_g \ge 5$, the precisions of StarGRUNet for $T_\texttt{eff}$ and $\log \ g$ are $94$ K and $0.16$ dex, respectively. The precisions of elemental abundances [X/H] are $0.07$ dex $\sim$ $0.16$ dex (except $0.18$ dex for [N/H] and $0.22$ dex for [Cr/H]). The test results show that StarGRUNet has higher accuracy and robustness compared with other catalog and neural networks on the whole.

To facilitate the use in astronomical science researches, this paper applied the trained StarGRUNet model to $8,208,332$ LAMOST-DR8 low-resolution spectra, computed the estimations for  $T_\texttt{eff}$, log~$g$, and 14 elements ([C/H], [Mg/H], [Al/H], [Si/H], [Ca /H], [Fe/H], [N/H], [O/H], [S/H], [Ti/H], [Cr/H], [Mn/H], [Ni/H], [K/H]). The estimates are also publicly released, and their URLs are available in the Acknowledgments section.

\section*{Acknowledgements}

This work is supported by the National Natural Science Foundation of China (Grant No. 11973022), the Natural Science Foundation of Guangdong Province (No. 2020A1515010710), the Major projects of the joint fund of Guangdong, and the National Natural Science Foundation (Grant No. U1811464). The authors are deeply grateful to Yu Lu, Jinqu Zhang, and Hui Li for their discussions in polishing this article.

LAMOST, a multi-target optical fiber spectroscopic telescope in the large sky area, is a major national engineering project built by the Chinese Academy of Sciences. Funding for the project is provided by the National Development and Reform Commission. LAMOST is operated and managed by the National Astronomical Observatory of the Chinese Academy of Sciences.

\section*{Data and code Availability}

The experimental dataset, estimated catalog, experimental code, and  trained models are available at \url{http://doi.org/10.12149/101216}.







\bibliographystyle{mnras}
\bibliography{cites}







\bsp	
\label{lastpage}
\end{document}